\documentclass[12pt]{article}
\usepackage{graphicx,epsf}
\newcommand{\be}{\begin{equation}}
\newcommand{\ee}{\end{equation}}
\newcommand{\ep}{\varepsilon}

\newcommand{\Li}[2]{{\mbox{Li}}_{#1}\left(#2\right)}
\newcommand{\Cl}[2]{{\mbox{Cl}}_{#1}\left(#2\right)}
\newcommand{\Ls}[2]{{\mbox{Ls}}_{#1}\left(#2\right)}

\newcommand{\tfrac}[2]{{\textstyle{\frac{#1}{#2}}}}

\newcommand{\Snp}[2]{{\mbox{S}}_{#1\!}\left(#2\right)}

\newcommand{\MSb}{$\overline{\rm MS}$ }
\newcommand{\MSbn}{$\overline{\rm MS}$}

\renewcommand{\thefootnote}{\fnsymbol{footnote}}

\begin{document}

\vspace*{-1cm}

\begin{flushright}
{\rm \normalsize%
{DESY 03-112}\\
{SFB/CPP-03-19}\\
{hep-ph/0308216} \\[3mm]
{October 20, 2003}
}
\end{flushright}

\vspace*{5mm}

\begin{center}
 {\large \bf
The $O(\alpha \alpha_s)$ correction to the pole mass of the t-quark within
the Standard Model }
 \end{center}
 \vspace{1cm}
 \begin{center}
{\large F.~Jegerlehner%
\footnote[1]{~E-mail: fred.jegerlehner@desy.de},
M.~Yu.~Kalmykov%
\footnote[2]{
Supported by DFG under Contract SFB/TR~9-03.
On leave from BLTP, JINR, 141980 Dubna, Russia.
E-mail: kalmykov@ifh.de}
}
\vspace{5mm}

{\it ~DESY Zeuthen, Platanenallee 6, D-15738, Zeuthen, Germany}

\vspace{5mm}
\end{center}
\begin{abstract}
We have calculated the $O(\alpha \alpha_s)$ contributions to the
relationship between the \MSbn--mass and the pole of the t-quark
propagator in the Standard Model in the limit of a diagonal CKM matrix
and for a massless b-quark.  Analytical results for the so far unknown
master--integrals appearing in the calculation are also given.
\end{abstract}
\renewcommand{\thefootnote}{\arabic{footnote}}
\setcounter{footnote}{0}

\section{Introduction}

Electroweak precision observables play an important role in the
verification of the Standard Model (SM) at the quantum level. By
comparing precision data with corresponding predictions it is possible
to get constraints on unknown parameters, like the Higgs boson mass,
or unveil new physics. For electroweak processes perturbative
calculations work and converge rather well. However, as the complexity
of such calculations grows dramatically with the order of the
perturbation expansion, complete higher order results are available in
a few cases only.  Therefore, perturbative calculations are possible
only at limited accuracy. We usually distinguish two types of
uncertainties: unknown higher order effects and the
experimental\footnote{In general they also include systematic
errors which stem from non--negligible theoretical uncertainties.}
errors of the input parameters.  For the calculation of electroweak
observables the generally accepted renormalization scheme, defining a
particular parametrization, is the so called {\em on-shell
scheme}~\cite{Sirlin80}--~\cite{Denner93}, where, in addition to the fine
structure constant (and/or the Fermi constant), the pole masses of
particles serve as input parameters.  Quark masses require special
consideration in this context, since on--shell quark masses are not
accessible experimentally.  Fortunately, the light quark masses in
high energy processes often can be neglected (effects $O(m_q^2/M_Z^2)$)
and thus can be treated as massless in practical calculations. The top
quark is different. The large numerical value of the top quark mass in
conjunction with the violation of the Appelquist-Carazzone
theorem~\cite{decoupling} as a consequence of the Higgs mechanism of
mass generation, implies that a class of radiative corrections are
proportional to positive powers of the top-quark mass which gives rise
to sizeable effects.  Moreover, the concept of a pole mass of a quark
is intrinsically ambiguous due to strong interaction renormalon
contributions~\cite{renormalon}.

The present experimental error of the top-quark mass, $\sim$ 5 GeV, will
be reduced to 1-2 GeV at the LHC and/or at a future Linear Collider.
The question about which type of mass will be determined actually with
this accuracy is discussed in~\cite{review}.  The higher order
theoretical uncertainties may be estimated from scheme-- and
scale--variations~\cite{scheme}.

The aim of the present paper is the analytical calculation of the
$O(\alpha \alpha_s)$ correction to the relationship between the pole--
and the \MSbn--mass of the top--quark. This also provides us the
two--loop on--shell mass counter-term for the top quark. The one--loop
results of order $O(\alpha_s)$ and $O(\alpha)$ have been presented
in~\cite{Tarrach} and e.g. in~\cite{BHS86}~\footnote{(see also Eq.~(B.5) 
in Appendix B of~\cite{poleII})}, respectively.  The
one-loop $O(\alpha)$ correction to the relationship between the
top-Yukawa coupling and the pole mass of the top-quark has been
calculated in~\cite{Yukawa}.  The two--loop $O(\alpha_s^2)$ correction
is given in~\cite{Broadhurst}, and the same order result obtained via
regularization by dimensional reduction may be found in~\cite{red}.
The renormalized off-shell fermion propagator of order $O(\alpha_s^2)$
has been worked out in~\cite{FJTV}. Only recently, in~\cite{pole3},
the three--loop $O(\alpha_s^3)$ correction has been
published. Finally, the two--loop $O(\alpha \alpha_s)$ and
$O(\alpha^2)$ corrections have been calculated in the approximation of
vanishing electroweak gauge couplings~\cite{rho3}. Our calculation,
presented here, extends previous two--loop $O(\alpha \alpha_s)$
calculations of the gauge boson self--energies~\cite{QCD} and the SM
$O(\alpha^2)$ corrections to the relation between the pole and the
\MSb masses of the gauge bosons $Z$ and $W$, presented
in~\cite{poleI,poleII}.

The paper is organized as follows: in Sec.~2 we outline the calculation
of the on--shell fermion self--energy. The relevant master--integrals
are presented in Sec.~3. In Sec.~4 we discuss the renormalization
problems at two--loops. Sec.~5 contains the main results of our
calculation and in Sec.~6 we summarize our conclusions.

\section{The quark pole mass: definition and calculation}

The definition of the top-quark pole mass has been discussed
in~\cite{toppole}. Starting point of our consideration is the tensor
decomposition of the one--particle irreducible self--energy of a
massive fermion $\tilde{\Sigma}(p, m,\ldots)$ which, within the SM,
has the form
\begin{eqnarray}
\hspace{-5mm}
\tilde{\Sigma}(p, m, \ldots) & \!\!=\!\! &
 i \hat{p} \left[ \tilde{A}(p^2,m,\ldots) \!-\! \gamma_5
 \tilde{C}(p^2,m,\ldots) \right] + m \left[ \tilde{B}(p^2,m,\ldots)
 \!-\! \gamma_5 \tilde{D}(p^2,m,\ldots) \right]\;,
\end{eqnarray}
where $\tilde{A}, \tilde{B}, \tilde{C}, \tilde{D}$ are Lorentz scalar
functions depending on all parameters of the SM.  For simplicity of
the notation we will indicate explicitly only the external momentum
$p^2$ and the mass $m$ of the quark under consideration.  The
$O(\alpha
\alpha_s)$ correction has a structure similar to the one in QCD, where
$\tilde{C} =\tilde{D} = 0$. In this case, the position of the pole
$-\tilde{M}$ is defined as the formal solution for $i \hat{p}$ (in
Euclidean metric \cite{diagramatic}
with $\hat{p}= \gamma p$, $\hat{p}^2=p^2$; on--shell:
$\hat{p}=im$, $p^2=-m^2$) at which the inverse of the connected full
propagator equals zero. Thus
\begin{equation}
i \hat{p} + m - \tilde{\Sigma}(p, m, \ldots)=i \hat{p}\:
\left(1-\tilde{A}(p^2,m,\ldots)\right) + m
\:\left(1-\tilde{B}(p^2,m,\ldots)\right) = 0\;
\label{pole}
\end{equation}
for $\hat{p}=i\tilde{M}$. The immediate question arising here is,
what is the interpretation of the complex mass $\tilde{M} \equiv
M'-\:\frac{i}{2}\: \Gamma'$ ? In general in a perturbative calculation
a transition amplitude $T$ exhibiting an unstable particle resonance
has the form ($U,V$ some spinor valued amplitudes)
\be
T=\bar{U}\: \frac{1}{i\hat{p}+\tilde{M}\:(1+O(p^2+\tilde{M}^2))}\:V 
=\frac{\bar{U}\:(-i\hat{p}+\tilde{M})\:V}{p^2+\tilde{M}^2+O((p^2+\tilde{M}^2)^2)}
=\frac{R}{p^2+\tilde{M}^2}+B(p^2)
\ee
where $R$ is a complex number and $B$ is a background complex scalar
amplitude which is regular at $p^2=-\tilde{M}^2$.  $T$ thus has the
same general form as in the case of a bosonic resonance. We thus
define the pole mass $M$ and the on--shell width $\Gamma$ as in the
bosonic case by
\be
\tilde{M}^2=M^2-iM\Gamma=M^{'2}-\Gamma^{'2}/4-iM'\Gamma'
\ee
such that
\be
M=\sqrt{M^{'2}-\Gamma^{'2}/4}\;\;;\;\;\;\Gamma=\frac{M'}{M}\Gamma'
\label{properMG}
\ee
Since $M=M' + O(\alpha^2)$ and $\Gamma=\Gamma' + O(\alpha^2)$ for the
$O(\alpha \alpha_s)$ terms considered in this paper we can identify
$M=M'$ and $\Gamma=\Gamma'$ in the following.

For the remainder of the paper we will adopt the following notation:
capital $M \simeq {\rm Re}\:\tilde{M}$ always denotes the pole mass;
lower case $m$ stands for the renormalized mass in the $\overline{\rm
MS}$ scheme, while $m_0$ denotes the bare mass. The on--shell width is
given by $\Gamma \simeq -2\: {\rm Im}\:\tilde{M}$. In addition we use
$e$, $g$ and $g_s$ to denote the $U(1)_{\rm em}$, $SU(2)_{\rm L}$ and
$SU(3)_{\rm c}$ couplings of the SM in the \MSb scheme.

In perturbation theory (\ref{pole}) is to be solved order by order.
For this aim we expand the self--energy function about the lowest
order solution $\hat{p}=i m_0$:
\begin{eqnarray}
\tilde{\Sigma}(p, m, \ldots) & = & \tilde{\Sigma} \left. \right|_{\hat{p}=i m_0}
+ \left(i \hat{p} + m_0  \right) \left. \left[\tilde{\Sigma}^{'}\right]
\right|_{\hat{p}=i m_0}
+ \cdots
\end{eqnarray}
and define dimensionless ``on--shell'' amplitudes $\Sigma$, $\Sigma'$
by
\begin{equation}
\left. \tilde{\Sigma} \right|_{\hat{p}=i m_0}
=\left. \left[-m_0 \tilde{A}+m_0 \tilde{B}\right]
\right|_{p^2=- m_0^2} \equiv -m_0 \Sigma (m_0,\ldots)
\label{sigma}
\end{equation}
and
\begin{equation}
\left. \left[\tilde{\Sigma}^{'}\right]
\right|_{\hat{p}=i m_0} = \left. \left[\left( \frac{\partial \tilde{\Sigma}}{\partial (i\hat{p})} \right) \right]
\right|_{\hat{p}=i m_0} =\left. \left[\tilde{A}+2p^2\dot{\tilde{A}}+2m_0^2\dot{\tilde{B}}\right]
\right|_{p^2=- m_0^2} \equiv \Sigma^{'}(m_0,\ldots)
\label{sigmaprime}
\end{equation}
where $\dot{X}(p^2,\ldots)$ denotes the derivative of $X(p^2,\ldots)$
with respect to $p^2$.  To two loops we then have the solution (in
agreement with~\cite{toppole})
\begin{eqnarray}
\hspace{-1cm}
\frac{\tilde{M}}{m} = 1 + \Sigma_1 + \Sigma_2 + \Sigma_1 \Sigma^{'}_1
\label{polemass}
\end{eqnarray} %
where $\Sigma_L$ is the bare ($m=m_0$) or \MSbn--renormalized ($m$ the
\MSbn--mass) $L$-loop contribution to the amplitudes defined in 
(\ref{sigma}) and (\ref{sigmaprime}).

\begin{figure}[th]
\centering
{\vbox{\epsfysize=120mm \epsfbox{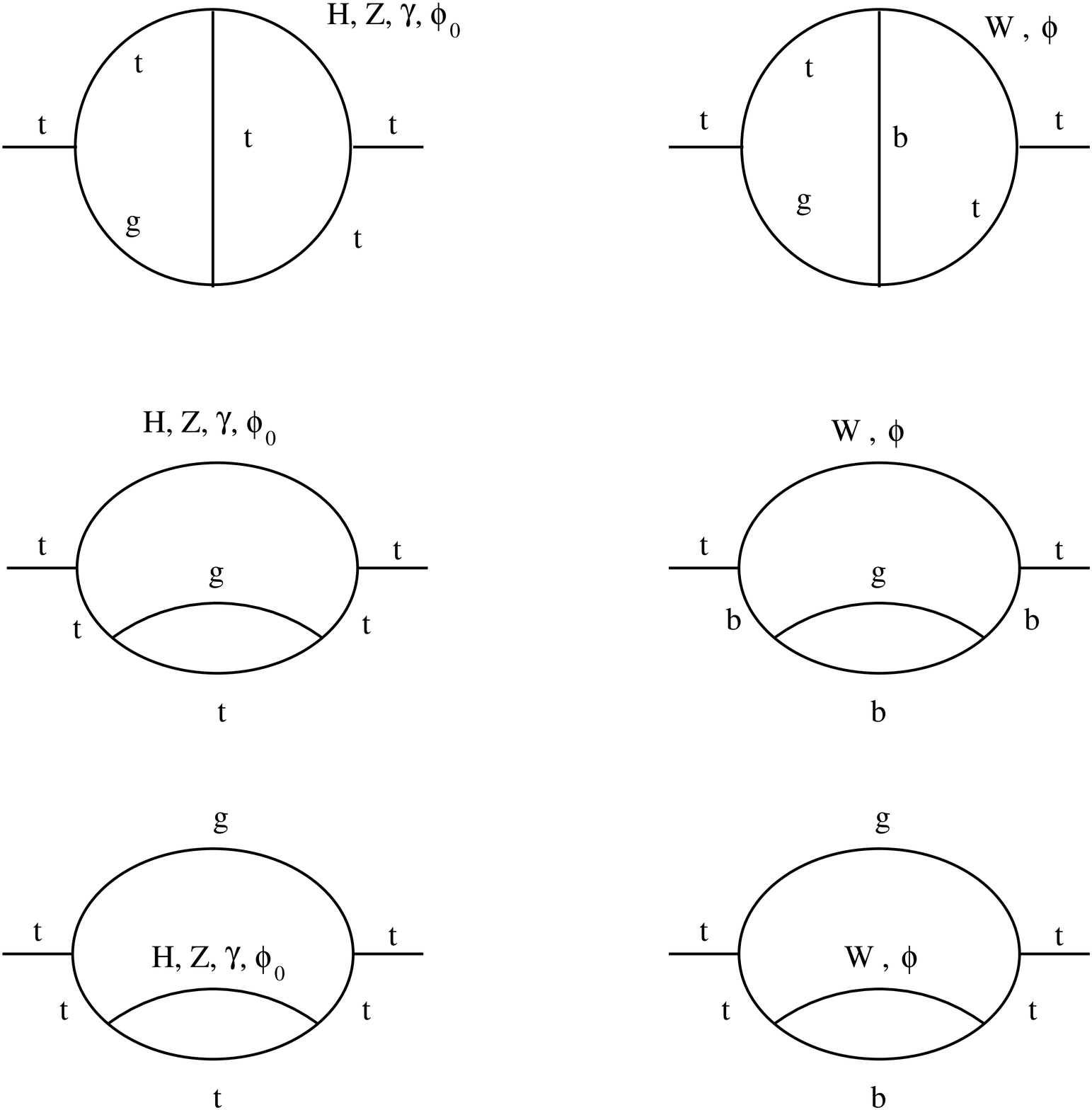}}}
\caption{
The two--loop one--particle irreducible diagrams contributing to the
pole mass of a quark.  $\phi_0$ and $\phi$ are the neutral and the
charged pseudo--Goldstone bosons, respectively. The number of diagrams
is 24.}
\label{diagrams2}
\end{figure}
\begin{figure}[th]
\centering
{\vbox{\epsfysize=60mm \epsfbox{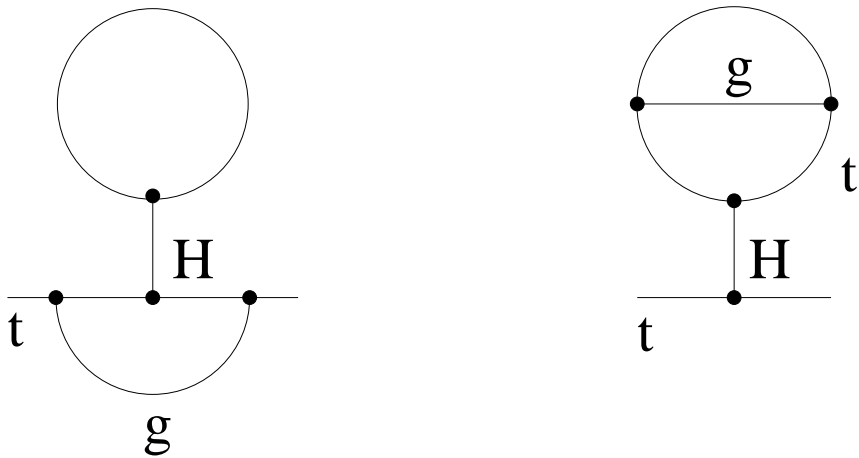}}}
\caption{
The two--loop tadpole type diagrams which should be included
for manifest gauge and renormalization group invariance.}
\label{tadpole}
\end{figure}

According to Eq.~(\ref{polemass}) we need to calculate propagator-type
diagrams up to two loops on--shell. The set of two--loop one--particle
irreducible diagrams is shown in Fig.~\ref{diagrams2}.  In order to
get manifestly gauge invariant results the Higgs tadpole diagrams,
shown in Fig.~\ref{tadpole} on the left side, should be included
\cite{FJ81}.  As was demonstrated in~\cite{poleII} for the
self--consistency of the renormalization group (RG) approach the
two--loop tadpoles of the type shown on the right side of
Fig.~\ref{tadpole} should be included as well.  For each quark
species, there is one $O(\alpha \alpha_s)$ two-loop 
tadpole diagram which is gauge invariant. For a quark with 
mass $m_q$ its bare contribution to the location of the pole of 
the top--quark propagator reads (see Sec.~4.3 in \cite{poleII})
\begin{eqnarray*}
\Delta_t = - \frac{g}{16 \pi^2 } \frac{g_s^2}{16 \pi^2}
16 N_c C_f \frac{1}{(d-4)^2} \frac{(d-1)}{(d-3)}
\frac{m_q^4}{m_W} (m_q^2)^{-2\ep} \;,
\end{eqnarray*}
with $C_f = 4/3$ in the SM and $N_c$ is the number of colors
($N_c=3$).  The one--loop ${\cal O }(\alpha)$ result for the pole mass
of a quark in the SM, in the approximation of a diagonal
Cabbibo-Kobayashi-Maskawa matrix is given by Eq.~(B.5) in Appendix~B
of~\cite{poleII}.  For the calculation of the two--loop propagator
type diagrams with several mass scales we will use Tarasov's
recurrence relations~\cite{T97a} which allow us to reduce all diagrams
to a few master--integrals. The package ${\bf
ONSHELL2}$~\cite{onshell2} is used for the calculation of the single
scale diagrams.

\section{Master--integrals}

\begin{figure}[th]
\centering
{\vbox{\epsfysize=40mm \epsfbox{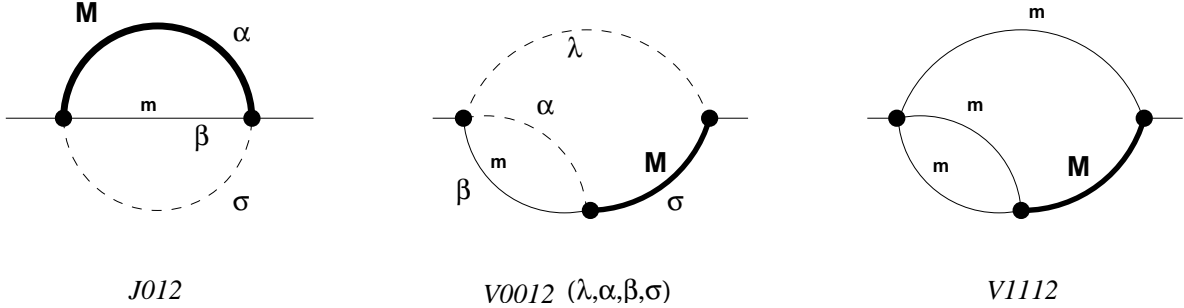}}}
\caption{
The master diagrams arising in this two--loop calculation.
Bold, thin and dashed lines correspond to off--shell massive,
on--shell massive and to massless propagators, respectively.}
\label{master}
\end{figure}

This section is devoted to the calculation of the so far unknown
two-loop master-integrals needed for our calculation and shown in
Fig.~\ref{master}.  We denote all master--integrals by
$T_{AB \cdots}$, where the first letter $T=F,V,J$ indicates the
topology in accordance with the notation introduced in~\cite{T97a};
indices $A,B,\cdots = 0,1,2$ characterize the relation between the
corresponding internal mass to the external momentum: $0$ indicates a
massless line, $1$ corresponds to ``internal mass equal to external
momentum'' and $2$ means that mass and momentum are different
(see~Fig.~\ref{master} for details). In our normalization each loop is
divided by $(4 \pi)^{2-\ep}\Gamma(1+\ep).$ We will also use the
short notations
\begin{eqnarray}
J_{m_1m_2m_3} & = & \left.
\frac{ \pi^{-n}}{\Gamma^2\left(3-\tfrac{n}{2}\right)}
\int \frac{d^n k_1 d^n k_2}{[(k_1-p)^2+m_1^2]
[k_2^2+m_2^2][(k_1-k_2)^2+m_3^3]}
\right|_{p^2=-m^2} \;,
\nonumber \\
A_0(M) & = & \frac{\pi^{-n/2}}{\Gamma\left(3-\tfrac{n}{2}\right)}
\int \frac{d^n k_1}{k_1^2+M^2}
\equiv \frac{4 M^{n-2}}{(n-2)(n-4)} \;.
\label{notations:integrals}
\end{eqnarray}
for the auxiliary integrals appearing in our calculation.

\subsection{$ J_{012}$}
The analytical result for this type of integral for arbitrary values of the
internal masses, momentum and powers of the propagators can be calculated by
using the Mellin--Barnes technique~\cite{BD-TMF}. It is expressible in 
terms of Appell hypergeometric function $F_4$ \cite{Appell}:
\begin{eqnarray}
&& \hspace{-5mm}
J_{012}(\sigma,\beta,\alpha,p^2,m^2,M^2) =
\int
\frac{d^nk_1 d^nk_2}{( (p-k_1)^2 )^{\sigma} ( (k_1-k_2)^2+M^2 )^{\alpha}
( k_2^2 + m^2)^{\beta}}
\nonumber \\ && \hspace{-5mm}
= (M^2)^{n-\sigma-\alpha-\beta}
\frac{\Gamma(\tfrac{n}{2}-\sigma)}{\Gamma(\sigma) \Gamma(\alpha) \Gamma(\beta)
 \Gamma(\tfrac{n}{2}) \Gamma^2(3-\tfrac{n}{2})}
\Biggl\{
\nonumber \\ && \hspace{-5mm}
\Gamma\left(\tfrac{n}{2} \!-\! \beta \right)
\Gamma\left(\alpha \!+\! \beta \!+\! \sigma \!-\! n \right)
\Gamma\left(\beta \!+\! \sigma \!-\! \tfrac{n}{2} \right)
F_4\left(\sigma\!+\!\beta\!-\!\tfrac{n}{2},
        \alpha \!+\! \beta \!+\! \sigma\!-\!n; \tfrac{n}{2}, 1\!+\!\beta\!-\!\tfrac{n}{2}
\left| \frac{-p^2}{M^2}, \frac{m^2}{M^2} \right. \right)
\nonumber \\ && \hspace{-5mm}
+ \left( \frac{m^2}{M^2} \right)^{(n/2-\beta)}
\Gamma\left(\beta \!-\! \tfrac{n}{2}\right)
\Gamma(\sigma)
\Gamma\left(\alpha \!+\! \sigma \!-\! \tfrac{n}{2}\right)
F_4\left(\sigma\!+\!\alpha\!-\!\tfrac{n}{2},
        \sigma; \tfrac{n}{2}, 1\!+\! \tfrac{n}{2} \!-\!\beta
\left| \frac{-p^2}{M^2}, \frac{m^2}{M^2} \right.\right)
\Biggr\}\;.
\end{eqnarray}
The result is symmetric with respect to the exchange $ M^2
\leftrightarrow m^2,\: \alpha \leftrightarrow \beta .  $ The
$\ep$-expansion for this diagram up to the finite part has been
given in~\cite{J012:F} and was recently recalculated
in~\cite{J012}.  In our case we need also the term linear in $\ep$,
however, only for the particular case, when the external momentum is
on-shell with respect to one of the internal masses.
In this case, taking $p^2 = - m^2$ (on--shell index $\beta$),
we simplify our notation and write
$$
J_{012}(\sigma,\beta,\alpha) =
\left. J_{012}(\sigma,\beta,\alpha,p^2,m^2,M^2) \right|_{p^2=-m^2}.
$$
We thus consider
\begin{eqnarray}
&& \hspace{-10mm}
J_{012}(\sigma,\beta,\alpha)=
(M^2)^{n-\sigma-\alpha-\beta}
\frac{\Gamma(\tfrac{n}{2}-\sigma)}
    {\Gamma(\sigma) \Gamma(\alpha) \Gamma(\beta)
\Gamma(\tfrac{n}{2}) \Gamma^2(3-\tfrac{n}{2})}
\Biggl\{
\nonumber \\ && \hspace{-10mm}
\Gamma\left(\tfrac{n}{2} \!-\! \beta \right)
\Gamma\left(\alpha \!+\! \beta \!+\! \sigma \!-\! n \right)
\Gamma\left(\beta \!+\! \sigma \!-\! \tfrac{n}{2} \right)
{}_{4}F_3 \left(\begin{array}{c|}
\alpha \!+\! \beta \!+\! \sigma \!-\! n,
\beta \!+\! \sigma \!-\! \tfrac{n}{2},
\tfrac{\beta}{2}, \tfrac{1+\beta}{2} \\
1 \!+\! \beta \!-\! \tfrac{n}{2},
\beta, \tfrac{n}{2}
\end{array} ~\frac{4m^2}{M^2} \right)
\nonumber \\ && \hspace{-10mm}
+ \left( \frac{m^2}{M^2} \right)^{(n/2-\beta)}
\Gamma\left(\beta \!-\! \tfrac{n}{2}\right)
\Gamma(\sigma)
\Gamma\left(\alpha \!+\! \sigma \!-\! \tfrac{n}{2}\right)
{}_{4}F_3 \left(\begin{array}{c|}
\sigma,
\alpha \!+\! \sigma \!-\! \tfrac{n}{2},
\tfrac{n-\beta}{2}, \tfrac{1+n-\beta}{2} \\
1 \!+\! \tfrac{n}{2} \!-\! \beta,
n \!-\! \beta, \tfrac{n}{2}
\end{array} ~\frac{4m^2}{M^2} \right)
\Biggr\}
\end{eqnarray}
in the following.  Let us remind that for the class of integrals
$J_{012}$ there are three master--integrals of this type:
$J_{012}(1,1,1)$, $J_{012}(1,2,1)$ and $J_{012}(1,1,2)$~\cite{T97a}.
However, other independent combinations happen to be more suitable for
performing the $\ep$-expansion: $J_{012}(1,2,2)$, $J_{012}(2,2,1)$, and
$[J_{012}(1,2,2)+J_{012}(2,1,2) + J_{012}(2,2,1)]$ (see also in Ref.~\cite{DS99}).  
The latter combination corresponds to the integral 
$J_{012}(1,1,1)$ in $2-2\ep$ dimensions~\cite{T96} and we have:
\begin{eqnarray}
J_{012}(1,2,2)
& = &
-\frac{(M^2)^{-1-2\ep}}{\ep(1-\ep)}
\Biggl\{
\frac{\Gamma(1-\ep) \Gamma(1+2\ep)}{\Gamma(1+\ep)}
{}_{3}F_2 \left(\begin{array}{c|}
1,\tfrac{3}{2}, 1+2\ep \\
2, 2-\ep
\end{array} ~\frac{4m^2}{M^2} \right)
\nonumber \\ && \hspace{25mm}
- \left( \frac{m^2}{M^2} \right)^{-\ep}
{}_{3}F_2 \left(\begin{array}{c|}
1,1+\ep,\tfrac{3}{2}-\ep \\
2-\ep,2-2\ep
\end{array} ~\frac{4m^2}{M^2} \right)
\Biggr\}
\;,
\end{eqnarray}

\begin{eqnarray}
&&
J_{012}(2,2,1)
=
\frac{(M^2)^{-1-2\ep}}{\ep^2(1-\ep)}
\Biggl\{
\frac{(1+\ep)\Gamma(1-\ep) \Gamma(1+2\ep)}{\Gamma(1+\ep)}
{}_{4}F_3 \left(\begin{array}{c|}
1,\tfrac{3}{2}, 1+2\ep, 2+\ep \\
2, 2-\ep, 1+\ep
\end{array} ~\frac{4m^2}{M^2} \right)
\nonumber \\ && \hspace{50mm}
- \left( \frac{m^2}{M^2} \right)^{-\ep}
{}_{3}F_2 \left(\begin{array}{c|}
2,1+\ep,\tfrac{3}{2}-\ep \\
2-\ep,2-2\ep
\end{array} ~\frac{4m^2}{M^2} \right)
\Biggr\}
\;,
\end{eqnarray}

\begin{eqnarray}
&&
J_c  \equiv J_{012}(1,2,2) + J_{012}(2,1,2) + J_{012}(2,2,1)
=
(M^2)^{-1-2\ep}
\frac{1}{\ep^2} \times
\nonumber \\ && \hspace{5mm}
\Biggl\{
\frac{\Gamma(1-\ep) \Gamma(2+2\ep)}{\Gamma(1+\ep)}
\Biggl[ 1 + 2 \frac{m^2}{M^2}  \frac{1+2\ep}{1-\ep}
{}_{3}F_2 \left(\begin{array}{c|}
1,\tfrac{3}{2}, 2+2\ep \\
2, 2-\ep
\end{array} ~\frac{4m^2}{M^2} \right)
\Biggr]
\nonumber \\ && \hspace{15mm}
- \left( \frac{m^2}{M^2} \right)^{-\ep}
\Biggl[ 1+ 2 \frac{m^2}{M^2} \frac{1+\ep}{1-\ep}
{}_{3}F_2 \left(\begin{array}{c|}
1,2+\ep,\tfrac{3}{2}-\ep \\
2-\ep,2-2\ep
\end{array} ~\frac{4m^2}{M^2} \right)
\Biggr]
\Biggr\}
\;.
\end{eqnarray}
The $\ep$-expansion for the given hypergeometric functions were worked
out in~\cite{poleII} and \cite{DK03}. While the individual expressions
are rather lengthy the results for the $\ep$-expansion of the integrals
have the compact form
\begin{eqnarray}
J_{012}(1,2,2)
& = &
(M^2)^{-1-2\ep} \frac{(1+y)^2}{y}
\Biggl\{
\ln (1+y) \Biggl[ \ln(1+y) - \ln(y) \Biggr]
\nonumber \\ && \hspace{-30mm}
+ \ep \Biggl[
2 \ln^3 (1+y)
- 2 \zeta_2 \ln (1+y)
-3 \ln y \ln^2 (1+y)
+ \frac{1}{2} \ln^2 y \ln (1+y)
\nonumber \\ && \hspace{-20mm}
- 2 \ln(y) \Li{2}{-y}
- 3 \ln (y) \Li{2}{y}
+ 6 \Li{3}{-y}
+ 6 \Li{3}{y}
\Biggr]
+ {\cal O}(\ep^2)
\Biggr\} \;,
\label{j122}
\end{eqnarray}

\begin{eqnarray}
&&
J_{012}(2,2,1)
= \frac{(M^2)^{-1-2\ep}}{(1-\ep)} \frac{(1+y)^2}{(1-y)}
\Biggl\{
\frac{1}{\ep} \Biggl[ \frac{1-y}{y} \ln (1 + y) + \ln y \Biggr]
- \frac{1}{2} \ln^2 y
+ 2 \zeta_2
\nonumber \\ && \hspace{5mm}
+
2 \frac{1-y}{y} \ln (1+y)
\Biggl(
\ln (1+y) - \ln y
\Biggr)
- \frac{1+y}{y}
\Biggl(
  3 \ln y \ln (1-y)
+ 4 \Li{2}{-y}
+ 3 \Li{2}{y}
\Biggr)
\nonumber \\ && \hspace{5mm}
+ \ep \Biggl[
\frac{1+y}{y}
\Biggl(
12 \Snp{1,2}{y^2}
- 4 \Snp{1,2}{-y}
- 6 \Snp{1,2}{y}
+ 24 \ln(1-y) \Li{2}{-y}
\nonumber \\ && \hspace{10mm}
+ 18 \ln(1-y) \Li{2}{y}
+ 9 \ln y \ln^2 (1-y)
+ \frac{3}{2} \ln^2 y \ln (1-y)
- 6 \zeta_2 \ln (1-y)
\Biggr)
\nonumber \\ && \hspace{10mm}
+ \frac{1-y}{y}
\Biggl(
\frac{8}{3} \ln^3 (1+y)
- 2 \zeta_2 \ln(1+y)
- 4 \ln y \ln ^2 (1+y)
+ \ln^2 y \ln (1+y)
\Biggr)
\nonumber \\ && \hspace{10mm}
+ \frac{1}{6} \ln^3 y
- 2 \zeta_3
+ \frac{15y + 9}{y} \ln y \Li{2}{y}
+ 4 \frac{3y + 2 }{y} \ln y \Li{2}{-y}
\nonumber \\ && \hspace{10mm}
- 4 \frac{7 y + 4}{y} \Li{3}{-y}
- 3 \frac{9 y + 5}{y} \Li{3}{y}
\Biggr]
+ {\cal O}(\ep^2)
\Biggr\}
\;,
\label{j221}
\end{eqnarray}

\begin{eqnarray}
&&
J_c  \equiv J_{012}(1,2,2) + J_{012}(2,1,2) + J_{012}(2,2,1)
\nonumber \\ &&
=
(M^2)^{-1-2\ep} \frac{1+y}{1-y}
\Biggl\{
\frac{1}{\ep} \ln y
- \Biggl[
 6 \ln y \ln (1-y)
+ \frac{1}{2} \ln^2 y
- 2 \zeta_2
+ 8 \Li{2}{-y}
+ 6 \Li{2}{y}
  \Biggr]
\nonumber \\ && \hspace{5mm}
+ \ep \Biggl[
  24 \Snp{1,2}{y^2}
- 8 \Snp{1,2}{-y}
- 12 \Snp{1,2}{y}
- 12 \zeta_2 \ln (1-y)
+ 48 \ln(1-y) \Li{2}{-y}
\nonumber \\ && \hspace{5mm}
+ 36 \ln (1-y) \Li{2}{y}
+ 18 \ln y \ln^2(1-y)
+ 3 \ln^2 y \ln(1-y)
+ \frac{1}{6} \ln^3 y
\nonumber \\ && \hspace{5mm}
+ 20 \ln y \Li{2}{-y}
+ 24 \ln y \Li{2}{y}
- 2 \zeta_3
- 44 \Li{3}{-y}
- 42 \Li{3}{y}
  \Biggr]
+ {\cal O}(\ep^2)
\Biggr\}
\;,
\label{jc}
\end{eqnarray}
where
\begin{equation}
y = \frac{1-\sqrt{1-4\frac{m^2}{M^2}}}{1+\sqrt{1-4\frac{m^2}{M^2}}} \;, \quad
\frac{M^2}{m^2} = \frac{(1+y)^2}{y} \;.
\label{y<->m}
\end{equation}

The expressions for the original set of master-integrals then read:
\begin{eqnarray}
&& \hspace{-5mm}
J_{012}(1,1,1,m^2,M^2) =
M^2 \frac{\left[ 4 m^2 - M^2 (9n-32) \right] }{(3n-10)(3n-8)(n-3)} J_c
\nonumber \\ && \hspace{5mm}
+  \frac{2 (M^2-m^2) \left[ 4 m^2 (2 n-7) + M^2 (n-4) \right] }{(3n-10)(3n-8)(n-3)}
\Biggl[ J_{012}(2,2,1) + J_{012}(1,2,2) \Biggr]
\nonumber \\ && \hspace{5mm}
- A_0(m) A_0(M) \frac{(n-2)^2 \left[(7n-24) M^2 + m^2(8n-28) \right]}{4 M^2 m^2 (3 n-10) (3n-8) (n-3))}
\; ,
\nonumber \\ && \hspace{-5mm}
J_{012}(1,2,1,m^2,M^2) =
- \frac{2 (M^2-m^2) (2 n-7)}{(3n-10)(n-3)} J_{012}(1,2,2)
- \frac{M^2}{(3n-10)(n-3)} J_c
\nonumber \\ && \hspace{5mm}
- \frac{\left[ M^2 (n-4) - 2 m^2 ( 2 n - 7) \right] }{(3n-10)(n-3)} J_{012}(2,2,1)
+ A_0(m) A_0(M) \frac{(n-2)^2(2n-7)}{4 M^2 m^2 (3 n-10) (n-3)}
\; ,
\nonumber \\ && \hspace{-5mm}
J_{012}(1,1,2,m^2,M^2) =
- \frac{\left[ 2 m^2 (n-3) + M^2 (n-4) \right] }{(3n-10)(n-3)}
       \Biggl[ J_{012}(2,2,1) + J_{012}(1,2,2) \Biggr]
\nonumber \\ && \hspace{5mm}
+ \frac{(3n-11) M^2}{(3n-10)(n-3)} J_c
+ A_0(m) A_0(M) \frac{(n-2)^2(2n-7)}{4 M^2 m^2 (3 n-10) (n-3)}
\; ,
\label{master_j}
\end{eqnarray}
where $A_0(m)$ is defined in (\ref{notations:integrals}).

\subsection{$V_{0012}$}
Let us consider the second diagram of Fig.~\ref{master},
\begin{eqnarray}
V_{00mM}(\lambda,\alpha,\beta,\sigma) =
\int
\frac{d^nk_1 d^nk_2}{( (p-k_1)^2 )^{\lambda} ( (k_1-k_2)^2 )^{\alpha} ( k_2^2 + m^2 )^{\beta}
                    ( k_1^2 + M^2)^{\sigma}}\;\;,
\end{eqnarray}
with the external momentum on the mass--shell, $p^2=-m^2.$ For this
class of integrals we have only one master-integral, the one with indices
$(1,1,1,1)$. For constructing the $\ep$-expansion we may use the
differential equation method \cite{kotikov}.  In terms of the new
variable $r = m^2/M^2$ the result may be represented as
\begin{eqnarray}
V_{00mM} = (m^2)^{-2\ep}
\left(
\frac{1}{2\ep^2} + \frac{S(r)}{\ep} + F(r) + \ep E(r) + {\cal O} (\ep^2)
\right) \; .
\end{eqnarray}
The analytical results for arbitrary $r$ read
\begin{eqnarray}
S(r) & = & \frac{5}{2} + \frac{(1-r)\ln(1-r)}{r} + \ln r \;,
\nonumber \\
F(r) & = & \frac{19}{2} - 2 \frac{1-r}{r} \Li{2}{r}
+ \zeta_2 (1-3r)
+ 4 \ln r
+ \left(1 - \frac{r}{2} \right) \ln^2 r
\nonumber \\ &&
+ \frac{(1-r)^2}{r} \ln r \ln(1-r)
+ \frac{4(1-r)}{r} \ln(1-r)
- \frac{(1-r)(3-r)}{2 r} \ln^2 (1-r) \; ,
\nonumber \\
E(r) & = &
\frac{65}{2}
- \frac{(1-r)(3r-7)}{r} \Snp{1,2}{r}
+ \frac{(1-r)(3r-5)}{r}  \Li{3}{r}
\nonumber \\ &&
+ \frac{2(1-r)(3-r)}{r} \Li{2}{r} \ln (1-r)
- \frac{1-r^2}{r} \Li{2}{r} \ln r
\nonumber \\ &&
- \frac{2(1-r)^2}{r} \ln r \ln^2 (1-r)
+ \frac{(1-r)(1-2r)}{r} \ln^2 r \ln(1-r)
\nonumber \\ &&
+ \frac{(4-3r)}{6} \ln^3 r
+ \frac{(1-r)(5-3r)}{3r} \ln^3(1-r)
- (1+3r) \zeta_3
\nonumber \\ &&
+ 2(1-r) \zeta_2 \ln r
- \frac{(1-r) (5r-3)}{r} \zeta_2 \ln (1-r)
- \frac{8(1-r)}{r}  \Li{2}{r}
\nonumber \\ &&
+ \frac{2(1-r)(r-3)}{r}\ln^2 (1-r)
+ \frac{4 (1-r)^2}{r}\ln r \ln(1-r)
+ 2(2-r) \ln^2 r
\nonumber \\ &&
+ 6(1-2r) \zeta_2
+ \frac{12(1-r)}{r} \ln (1-r)
+ 12 \ln r \; .
\label{v00mM}
\end{eqnarray}
For $r=1$ (on--shell case) we have: $$ S(1) = \frac{5}{2} \;, \quad
F(1) = \frac{19}{2} -2
\zeta_2 \;, \quad E(1) = \frac{65}{2} - 6 \zeta_2 - 4 \zeta_3 \; .  $$

\subsection{$V_{1112}$}
The last and most complicated master--integral is
\begin{equation}
V_{mmmM}(\alpha,\beta,\sigma,\lambda) =
\int
\frac{d^nk_1 d^nk_2}{( (p-k_1)^2 + m^2)^{\sigma} ( (k_1-k_2)^2 + m^2)^{\alpha} ( k_2^2 + m^2 )^{\beta}
                    ( k_1^2 + M^2)^{\lambda}}
\end{equation}
where the external momentum is on the mass--shell, $p^2=-m^2.$ Let us
introduce here an angle $\theta$ defined via \cite{DD98}
\begin{eqnarray}
\cos\theta = \frac{M}{2m} \;, M \leq 2m \;.
\end{eqnarray}
The $\ep$-expansion now can be written as
\begin{eqnarray}
V_{mmmM}(\theta) = (m^2)^{-2 \ep} \Biggl[
\frac{1}{2 \ep^2}
+ \frac{1}{\ep } S(\theta)
+ F(\theta)
+ \ep E(\theta)
+ {\cal O}(\ep^2)
\Biggr] \;.
\end{eqnarray}
The analytical results for arbitrary values of the angle $\theta$ read
\begin{eqnarray}
S(\theta) & = &
 \frac{5}{2}
- 4 \theta \cos \theta \sin \theta
- 4 \cos^2 \theta \ln \left( 2 \cos \theta \right) \; ,
\nonumber \\
F(\theta) & = &
\frac{19}{2}
+ 4 \theta \sin (2 \theta )\ln \left( 2 \sin 2 \theta \right)
- 8 \theta \sin (2 \theta )
- 4 \theta^2 \sin^2 \theta
+ 2 \zeta_2  \sin^2 \theta
\nonumber \\ &&
- \sin (2 \theta ) \Biggl[ 4 \Ls{2}{2\theta} - 3 \Ls{2}{4\theta} \Biggr]
+ 4 \cos^2 \theta \ln^2 \left( 2 \cos \theta \right)
- 16 \cos^2 \theta \ln \left( 2 \cos \theta \right) \; ,
\nonumber \\
E(\theta) & = &
\frac{65}{2}
 - 24 \theta \sin (2 \theta)
 - 16 \theta^2 \sin^2 \theta
 + 12 \Cl{3}{2 \theta} \sin^2 \theta
 - 3 \Cl{3}{4 \theta} \sin^2 \theta
\nonumber \\ &&
 + 16 \theta \sin ( 2 \theta)  \ln \left( 2 \sin (2 \theta) \right)
 + 16 \theta^2 \sin^2  \theta  \ln \left( 2 \sin \theta \right)
 - 4 \theta \sin ( 2 \theta) \ln^2 \left( 2 \sin (2 \theta) \right)
\nonumber \\ &&
 + 8  \sin ( 2 \theta) \ln \left( 2 \sin (2 \theta) \right) \Ls{2}{2\theta}
 - 6  \sin ( 2 \theta) \ln \left( 2 \sin (2 \theta) \right) \Ls{2}{4\theta}
\nonumber \\ &&
 - 48 \ln \left( 2 \cos \theta \right)
 + 8 \theta^2 \sin^2  \theta \ln \left( 2 \cos \theta \right)
 + 48 \sin^2  \theta \ln \left( 2 \cos \theta \right)
\nonumber \\ &&
 + 16 \cos^2  \theta \ln^2 \left( 2 \cos \theta \right)
 - \frac{8}{3} \cos^2 \theta \ln^3 \left( 2 \cos \theta \right)
 + 8 \theta \sin^2  \theta \Ls{2}{2\theta}
\nonumber \\ &&
 - 16 \sin ( 2 \theta) \Ls{2}{2\theta}
 + 12 \sin ( 2 \theta) \Ls{2}{4\theta}
 - 4 \sin ( 2 \theta)  \Ls{3}{2\theta}
 + 3 \sin ( 2 \theta)  \Ls{3}{4\theta}
\nonumber \\ &&
 - 12 \zeta_2 \sin^2  \theta \ln \left( 2 \cos \theta \right)
 - 4 \zeta_2
 + 8 \zeta_2  \sin^2  \theta
 - 2 \zeta_3 \sin^2  \theta
\;,
\label{VmmmM:angle}
\end{eqnarray}
where the $\Ls{j}{\theta}$ are so-called log-sine integrals~\cite{Lewin} defined by
\begin{equation}
\Ls{j}{\theta} =   - \int\limits_0^\theta {\rm d}\phi \;
                    \ln^{j-1} \left| 2\sin\frac{\phi}{2}\right| \; ,
\label{log-sin}
\end{equation}
$\Cl{j}{\theta}$ are Clausen functions,

\begin{eqnarray}
\Cl{j}{\theta}=
\left\{ \begin{array}{l}
\!\!\! {\textstyle{1\over2{\rm i}}}
\left[ {\mbox{Li}}_j\left(e^{{\rm i}\theta}\right)      \!-\! {\mbox{Li}}_j\left(e^{-{\rm i}\theta}\right) \right] ,
\hspace{1mm} j \; \mbox{even} \\[2mm]
\!\! {\textstyle{1\over2}}
\left[ {\mbox{Li}}_j\left(e^{{\rm i}\theta}\right)      \!+\! {\mbox{Li}}_j\left(e^{-{\rm i}\theta}\right) \right] ,
\hspace{1mm} j \; \mbox{odd}
\end{array}\!\!\! \right.
\end{eqnarray}
and the $\mbox{Li}_j$ are poly-logarithms.

For $M^2=m^2$ the angle $\theta$ is equal to $\pi/3$ and the integral
is equal to the single scale diagram ${\bf V1111}$ calculated
in~\cite{single}.  In this case we obtain
\begin{eqnarray}
S\left(\frac{\pi}{3} \right) & = & \frac{5}{2} - \frac{\pi}{\sqrt{3}} \;,
\quad
F\left(\frac{\pi}{3}  \right) = \frac{19}{2}
+ \frac{\pi}{\sqrt{3}} \ln 3
- 4  \frac{\pi}{\sqrt{3}} - \frac{1}{2} \zeta_2
- 7 \frac{\Ls{2}{\frac{\pi}{3}}}{\sqrt{3}} \;,
\nonumber \\
E\left(\frac{\pi}{3}  \right) & = &
\frac{65}{2}
- 6 \zeta_2
+ 4 \frac{\pi}{\sqrt{3}} \ln 3
- \frac{1}{2} \frac{\pi}{\sqrt{3}} \ln^2 3
- 12 \frac{\pi}{\sqrt{3} }
- \frac{9}{2} \frac{\pi}{\sqrt{3}} \zeta_2
+ 4 \zeta_2 \ln 3
\nonumber \\ &&
- \frac{9}{2} \zeta_3
+ 7 \frac{\Ls{2}{\frac{\pi}{3}}}{\sqrt{3}} \ln 3
- 28 \frac{\Ls{2}{\frac{\pi}{3}}}{\sqrt{3}}
+ \frac{4}{3} \pi \Ls{2}{\frac{\pi}{3}}
- \frac{21}{2}\frac{\Ls{3}{\frac{2 \pi}{3}}}{\sqrt{3}}  \;.
\end{eqnarray}
Another case of interest is the massless case $M^2=0$, which corresponds to
$\theta = \pi/2$. In this case the diagram can  be reduced to
$J_{mmm}(1,1,1)$ \cite{ON3A,ON3B} plus a combination of simpler vacuum 
diagrams \cite{DT}. The results here are
$$
S\left(\frac{\pi}{2}  \right) = \frac{5}{2} \;, \quad
F\left(\frac{\pi}{2}  \right) =  \frac{19}{2} - 4 \zeta_2 \;,  \quad
E\left(\frac{\pi}{2}  \right) =
\frac{65}{2}
+ 24 \zeta_2 \ln 2
- 20 \zeta_2
- 14 \zeta_3 \;.
$$
The expression (\ref{VmmmM:angle}) is directly applicable in the region
$M \leq 2m$ only.  For $(M > 2m)$ one has to perform the proper analytical
continuation. How this can be done is described in details
in~\cite{DK-bastei,DK01,DK03}. For this purpose, let us introduce the
new variable
\begin{eqnarray}
y \equiv e^{ {\rm i} \sigma 2 \theta}, \hspace{5mm}
\ln(-y-{\rm i}\sigma 0) = \ln{y} - {\rm i} \sigma \pi.
\end{eqnarray}
which coincides with the variable $y$ defined in~(\ref{y<->m}).
In terms of this variable the expression (\ref{VmmmM:angle}) may be written as
\begin{eqnarray}
S(y) & = &
\frac{5}{2} + (1+y) \ln y - \frac{(1+y)^2}{y} \ln (1+y) \;,
\nonumber \\
F(y) & = &
\frac{19}{2}
+ \frac{1-y^2}{y} \Biggl[ \Li{2}{y} + 3 \Li{2}{-y} \Biggr]
+ \ln^2 y
+ (1-y) \zeta_2
\nonumber \\ &&
+ (1+y) \Biggl(
4 \ln y
- 2 \ln y \ln (1-y)
- 2 \ln y \ln (1+y)
         \Biggr)
\nonumber \\ &&
+ \frac{(1+y)^2}{y} \Biggl[
\ln^2 (1+y)
- 4  \ln (1+y)
+ \ln y \ln (1-y)
\Biggr]
\; ,
\nonumber \\
E(y) & = &
\frac{65}{2}
+ \frac{1-y^2}{y} \Biggl[
  4\Snp{1,2}{y}
- 3 \Snp{1,2}{y^2}
- 6 \ln (1-y) \Li{2}{-y}
- 2 \ln (1-y) \Li{2}{y}
\nonumber \\ &&
+ 4 \Li{3}{y}
- 6 \ln (1+y) \Li{2}{-y}
- 2 \ln (1+y) \Li{2}{y}
- 2 \ln y \ln (1-y) \ln (1+y)
\nonumber \\ &&
- \ln y \ln^2 (1-y)
- \zeta_2 \ln (1-y)
+ 12 \Li{2}{-y}
+ 4 \Li{2}{y}
+ 4 \ln y \ln (1-y)
                 \Biggr]
\nonumber \\ &&
- 12 \zeta_2 \ln (1+y)
+ 4 \ln^2 y
+ 2 \frac{(1+2y)(1+y)}{y} \zeta_2 \ln (1+y)
- 4 \ln^2 y \ln (1-y)
\nonumber \\ &&
- 2 \ln^2 y \ln (1+y)
+ \ln^3 y
+ (1-y) ( 2 \zeta_2 \ln y  - \zeta_3 )
+ 4 y \ln y \Li{2}{y}
- 4 y \zeta_2
\nonumber \\ &&
+ (1+y) \Biggl[
  12 \ln y
- 8 \ln y \ln (1+y)
+ 2 \ln y \ln^2 (1+y)
- \frac{1}{3} \ln^3 y
\Biggr]
\nonumber \\ &&
+ \frac{(1+y)^2}{y} \Biggl[
4 \ln^2 (1+y)
- 12 \ln (1+y)
- \frac{2}{3} \ln^3 (1+y)
+ \ln^2 y \ln (1-y)
\nonumber \\ &&
- \ln y \Li{2}{y}
\Biggr]
- \frac{3}{y} (y-1)(y+3) \Li{3}{-y} \; .
\label{VmmmM:y}
\end{eqnarray}

The results (\ref{j122})-(\ref{master_j}),(\ref{v00mM}) (\ref{VmmmM:y})
have been checked by a heavy mass expansion~\cite{asymptotic} with the
help of the packages described in~\cite{tlamm}.

%
\section{Renormalization}

The mass renormalization constant $Z_t$ in the \MSb scheme at two
loops may be written in the form
\begin{eqnarray}
m_{t,0} & = &  m_t(\mu^2)\;Z_t
= m_t(\mu^2)\:
\Biggl( 1
+ \frac{g^2(\mu^2)}{16\pi^2}\;\frac{1}{\ep} Z_\alpha^{(1,1)}
+ \frac{\alpha_s(\mu^2)}{4 \pi}\;\frac{1}{\ep} Z_{\alpha_s}^{(1,1)}
\nonumber \\ && \hspace{5mm}
+ \frac{\alpha_s(\mu^2)}{4 \pi } \frac{g^2(\mu^2)}{16\pi^2}
\left( \frac{1}{\ep} Z_{\alpha \alpha_s}^{(2,1)} +
\frac{1}{\ep^2} Z_{\alpha \alpha_s}^{(2,2)}
\right)
+ {\cal O}(g^4, \alpha_s^2)
\Biggr),
\label{baremsb}
\end{eqnarray}
where $\alpha_s = g_s^2/4 \pi$. Note that in contrast to the
corresponding definition for bosons (used, e.g.,
in~\cite{poleI,poleII}), the mass renormalization constant here (i.e.,
for fermions) relates the masses and not the squared ones. The
coefficient $Z_\alpha^{(1,1)}$ may be extracted from Eq.(4.31)
of~\cite{poleII} ($ 2 Z_\alpha^{(1,1)} = Z^{(1,1)}_{\rm t-quark}$).
The coefficient $Z_{\alpha_s}^{(1,1)}$ is well known
\cite{Tarrach}. For completeness we give them here:

\begin{eqnarray}
Z_\alpha^{(1,1)} & = &
\frac{1}{3}
- \frac{1}{3} \frac{m_Z^2}{m_W^2}
- \frac{3}{4} \frac{m_Z^4}{m_W^2 m_H^2}
- \frac{3}{8} \frac{m_H^2}{m_W^2}
- \frac{3}{2} \frac{m_W^2}{m_H^2}
+ \frac{3}{8} \frac{m_t^2}{m_W^2}
+ N_c \frac{m_t^4}{m_W^2 m_H^2} \;,
\nonumber \\
Z_{\alpha_s}^{(1,1)}  & = & - 3 C_f.
\label{RG:one-loop}
\end{eqnarray}
In our calculation we obtained the two-loop renormalization constants
$Z_{\alpha \alpha_s}^{(2,1)}$ and $Z_{\alpha \alpha_s}^{(2,2)}$
\begin{eqnarray}
&&
Z_{\alpha \alpha_s}^{(2,2)} = C_f \Biggl[
\frac{m_Z^2}{m_W^2}
+ \frac{9}{4} \frac{m_Z^4}{m_W^2 m_H^2}
- 9 N_c \frac{m_t^4}{m_H^2 m_W^2}
+ \frac{9}{8} \frac{m_H^2}{m_W^2}
- \frac{9}{4} \frac{m_t^2}{m_W^2}
+ \frac{9}{2} \frac{m_W^2}{m_H^2}
- 1
\Biggr] \;,
\label{Z22}
\\ &&
Z_{\alpha \alpha_s}^{(2,1)} =
 C_f \Biggl[
2 N_c \frac{m_t^4}{m_W^2 m_H^2}
+ \frac{3}{2} \frac{m_t^2}{m_W^2}
+ \frac{19}{48} \frac{m_Z^2}{m_W^2}
+ \frac{31}{24}
\Biggr] \;,
\label{Z21}
\end{eqnarray}
where, in the SM, $C_f = 4/3$, $N_c = 3$ and the first five quarks are
treated as massless.  The terms proportional to $m_t^4$ are coming from
the tadpole contribution and will cancel in observable quantities.  We
may use the SM renormalization group equations to cross-check the
$1/\ep^2$-- and $1/\ep$--terms (see also~\cite{poleI,poleII}). The
coefficient (\ref{Z22}) of the $1/\ep^2$ term may be calculated from the
relations
\begin{eqnarray}
\left(
\gamma_t +\sum_j \beta_{g_j} \frac{\partial}{\partial g_j }
+ \sum_i \left[ \mu^2 \frac{\partial}{\partial \mu^2} m_i^2(\mu^2) \right]
\frac{\partial}{\partial m_i^2} \right) Z_t^{(1)}
& = &  \frac{1}{2} \sum_j g_j \frac{\partial}{\partial g_j } Z_t^{(2)},
\label{RG:SM+QCD}
\end{eqnarray}
where we adopted the notation $m_{t,0} = m_t \left(1 + \sum_k
Z^{(k)}_t/\ep^k \right)$ and $g_j = g, g_s$.  The anomalous dimension
of the top-quark mass $\gamma_t$ is defined by
\begin{equation}
\gamma_t \equiv \frac{1}{m_t} \mu^2 \frac{\partial}{\partial \mu^2} m_t(\mu^2)
=  \frac{1}{2} \sum_j g_j \frac{\partial}{\partial g_j } Z_t^{(1)} \;.
\end{equation}
Translated into a relation for the coefficients $Z^{(i,j)}$ defined
in~(\ref{baremsb}) we have
\begin{eqnarray}
\gamma_t & = &
\frac{1}{16 \pi^2} \Biggl(
g^2 Z_\alpha^{(1,1)} + g_s^2 Z_{\alpha_s}^{(1,1)}
                           \Biggr)
+ \frac{g^2 g_s^2}{ (16 \pi^2)^2} 2 Z_{\alpha \alpha_s}^{(2,1)}
+ {\cal O}(g^4,\alpha_s^2) \;,
\nonumber \\
Z_{\alpha \alpha_s}^{(2,2)}  & = &
Z_{\alpha_s}^{(1,1)}
\left( 1 + m_t^2 \frac{\partial}{\partial m_t^2} \right) Z_\alpha^{(1,1)}
= Z_{\alpha_s}^{(1,1)}
\left(Z_\alpha^{(1,1)} + \frac{3}{8}
\frac{m_t^2}{m_W^2} + 2 N_c \frac{m_t^4}{m_W^2 m_H^2} \right)
\:.
\label{QCD}
\end{eqnarray}
Note that (\ref{QCD}) explicitly reveals, that the systematic
inclusion of the tadpole contributions is important for the
self--consistency of the RG equations.

The terms proportional to $1/\ep$ may be deduced from the RG equations
calculated in the unbroken phase.  It has been shown
in~\cite{poleI,poleII} (details are given in~\cite{expansion}) that in
the \MSb scheme we may write
\begin{equation}
m_t^2 (\mu^2) = \frac{1}{2} \frac{Y^2_t(\mu^2)}{\lambda(\mu^2)}\;
m^2(\mu^2) \;,
\end{equation}
where $m^2$ and $\lambda$ are the parameters of the symmetric scalar
potential and $Y_t$ is the top-quark Yukawa coupling. As a consequence
we get the following relation for the anomalous dimension of the mass
of the top-quark $\gamma_t$
\begin{equation}
\gamma_t = \gamma_{Y} + \frac{1}{2} \gamma_{m^2} 
- \frac{1}{2} \frac{\beta_\lambda}{\lambda} \;,
\label{SM<->F}
\end{equation}
where the relevant RG results may be found in~\cite{RG_2loop}:
\begin{eqnarray}
&&
\gamma_{m^2} \equiv
\frac{1}{m^2} \mu^2 \frac{\partial}{\partial \mu^2} m^2  =
\frac{1}{16 \pi^2}
\Biggl[ \lambda + 3 Y_t^2 - \frac{9}{4} g^2 - \frac{3}{4} {g'}^2
\Biggr]
+ 20  \frac{Y_t^2 g_s^2 }{(16 \pi^2)^2}
+ {\cal O} (g^4)
\; ,
\nonumber \\
&&
\gamma_{Y} \equiv
\frac{1}{Y_t} \mu^2 \frac{\partial}{\partial \mu^2} Y_t  =
\frac{1}{16 \pi^2}
\Biggl[ \frac{9}{4} Y_t^2 - 4 g_s^2 - \frac{9}{8} g^2 - \frac{17}{24} {g'}^2
\Biggr]
\nonumber \\ &&  \hspace{35mm}
+
\frac{g_s^2}{(16 \pi^2)^2}
\Biggl[
18 Y_t^2
+ \frac{9}{2} g^2
+ \frac{19}{18} {g'}^2
\Biggr]
+ {\cal O} (g^4)
\;,
\nonumber \\
&&
\beta_\lambda \equiv \mu^2 \frac{\partial}{\partial \mu^2} \lambda
= \frac{1}{(16 \pi^2)}
\Biggl[
2 \lambda^2
+ 6 \lambda Y_t^2
- 18 Y_t^4
- \frac{9}{2} \lambda g^2
- \frac{3}{2} \lambda g'^2
+ \frac{27}{8} g^4
+ \frac{9}{4} g^2 g'^2
+ \frac{9}{8} g'^4
\Biggr]
\nonumber \\ &&  \hspace{35mm}
+ \frac{g_s^2 Y_t^2 }{(16 \pi^2)^2}
\Biggl[
40 \lambda - 96 Y_t^2
\Biggr]
+ {\cal O} (g^4)
\;.
\label{RG}
\end{eqnarray}
Finally, the parameter relations
\begin{eqnarray*}
Y_t^2 = \frac{2 m_t^2}{v^2} \;, \quad
\lambda = \frac{3 m_H^2}{v^2} \;, \quad
g^2 = \frac{4 m_W^2}{v^2} \;, \quad g'^2 = \frac{4 (m_Z^2-m_W^2)}{v^2}
\;, \quad
\end{eqnarray*}
provide the necessary bridge between Eqs.~(\ref{SM<->F}) and
(\ref{RG}) and our Eqs.~(\ref{RG:one-loop}) and (\ref{Z21}).

We now turn to the discussion of the pole--mass relation
(\ref{polemass}).  The calculation of the one-loop \MSb renormalized
on--shell amplitude $\Sigma_1$ (see (\ref{sigma}) for the definition)
is simple.  We get it by rewriting the bare expression in terms of
\MSb parameters. In terms of the amplitudes $X$ defined by
\begin{equation}
\Sigma_{1,0}= \frac{g_{s,0}^2}{16\pi^2} X^{(1)}_{\alpha_s,0}
+  \frac{g_0^2}{16\pi^2} X^{(1)}_{\alpha,0}
\end{equation}
we obtain
\begin{eqnarray}
\Biggl \{ \Sigma_1 \Biggr\}_{\overline{\rm MS}}
& \equiv &
\lim_{\ep \to 0} \left( \frac{m_{0,t}}{m_t(\mu)}
\Biggl[ 1 + \frac{g_{s,0}^2}{16\pi^2} X^{(1)}_{\alpha_s,0}
+  \frac{g_0^2}{16\pi^2} X^{(1)}_{\alpha,0}
\Biggr]
- 1
\right)
\nonumber \\
& = & \lim_{\ep \to 0}
\left(
\frac{\alpha_s}{4 \pi}\; \Biggl[ X^{(1)}_{\alpha_s,0}
+ \frac{1}{\ep} Z_{\alpha_s}^{(1,1)}\Biggr]
+ \frac{e^2}{16\pi^2 \sin^2 \theta_W} \Biggl[ X^{(1)}_{\alpha,0} +
\; \frac{1}{\ep} Z_\alpha^{(1,1)}  \Biggr] \right)
\nonumber \\
& = &
\frac{\alpha_s}{4\pi}
\; X_{\alpha_s}^{(1)}
+ \frac{e^2}{16\pi^2 \sin^2 \theta_W}
\; X_{\alpha}^{(1)}
\;,
\label{MS1:subtracted}
\end{eqnarray}
where $X_{\alpha_s}^{(1)} = 4 C_f + Z^{(1,1)}_{\alpha_s} \ln
\frac{m_t^2}{\mu^2}$ and $X_{\alpha}^{(1)}=\frac{1}{2} X_{top}^{(1)}$
where $X_{top}^{(1)}$ is given in Eq.~(B.5) of Ref.~\cite{poleII}.  We
note that the explicit $\mu$--dependence is given by the following
structure:
\begin{equation}
X_{i}^{(1)} =\Delta X_{i}^{(1)} + Z^{(1,1)}_{i} \ln \frac{m_t^2}{\mu^2} \;\;;
\;\;(i=\alpha,\alpha_s)
\end{equation}
where $\Delta X_{i}^{(1)}$ does not explicitly depend on $\mu$.
At the two-loop level, we may avoid
the consideration of the wave-function renormalization as well as the
renormalization of the ghost sector and of the gauge parameters if we
look directly at the full two-loop \MSb renormalized on--shell
amplitude.  The latter can be written in the form

\begin{eqnarray}
&&
\Biggl \{ \Sigma_2 + \Sigma_1 \Sigma_1^{'} \Biggr\}_{\overline{\rm MS}}
=  \lim_{\ep \to 0}  \Biggl(
\Sigma_{2,0} + \Sigma_{1,0} \Sigma_{1,0}^{'}
+
\frac{\alpha_s}{4 \pi}
\frac{e^2}{16\pi^2 \sin^2 \theta_W}
\Biggl[ \frac{1}{\ep} Z_{\alpha \alpha_s}^{(2,1)}
+ \frac{1}{\ep^2} Z_{\alpha \alpha_s}^{(2,2)}
\Biggr]
\nonumber \\ &&
+
\frac{\alpha_s}{4 \pi} \frac{e^2}{16\pi^2 \sin^2 \theta_W}\:\frac{1}{\ep}\:
\Biggl\{
Z_{\alpha}^{(1,1)}
\Biggl[ 1 + 2 m_t^2 \frac{\partial}{\partial m_{t,0}^2} \Biggr] X_{\alpha_s,0}^{(1)}
+  Z_{\alpha_s}^{(1,1)}
\Biggl[ 1 + 2 m_t^2 \frac{\partial}{\partial m_{t,0}^2} \Biggr] X_{\alpha,0}^{(1)}
\Biggr\} \Biggr)
\nonumber \\ &&
= \frac{\alpha_s}{4 \pi} \frac{e^2}{16\pi^2 \sin^2 \theta_W}
\Biggl(
C^{(2,2)}_{\alpha \alpha_s} \ln^2 \frac{m_t^2}{\mu^2}
+  C^{(2,1)}_{\alpha \alpha_s} \ln  \frac{m_t^2}{\mu^2}
\nonumber \\ &&
+ \frac{1}{8} \ln \left( 1 - \frac{1}{\omega_t} \right) \left( 1 - \omega_t \right)
\frac{( 18 \omega_t^2 + 21 \omega_t + 17)}{\omega_t}
+ \ln \omega_t \Biggl[ \frac{22 \omega_t + 17}{8\omega_t} \Biggr]
\nonumber \\ &&
+ \left( 1 - \omega_t \right) \frac{(1 + 2 \omega_t)(2 + \omega_t)}{2 \omega_t}
\ln \omega_t  \ln \left( 1 - \frac{1}{\omega_t} \right)
+ \frac{1 + \omega_t - \omega_t^2}{2 \omega_t} \ln^2 \omega_t
\nonumber \\ &&
- (1-\omega_t)^2 \frac{4 \omega_t + 5}{8 \omega_t}
  \ln^2 \left( 1 - \frac{1}{\omega_t} \right)
+ \frac{(1+\omega_t)}{4 \omega_t} \Biggl( 4 \omega_t^2  + 7 \omega_t - 9 \Biggr)
  \Li{2}{\frac{1}{\omega_t}}
\nonumber \\ &&
- \frac{1}{\omega_t} (1-\omega_t)^2 (1 + 2 \omega_t)
\Biggl \{
  \frac{3}{2} \Snp{1,2}{\frac{1}{\omega_t}}
- \frac{3}{2} \Li{3}{\frac{1}{\omega_t}}
- \ln \omega_t  \Li{2}{\frac{1}{\omega_t}}
\nonumber \\ &&
+ \frac{1}{2}  \ln \left( 1 - \frac{1}{\omega_t} \right)  \Li{2}{\frac{1}{\omega_t}}
+ \frac{1}{4} \ln \left( 1 - \frac{1}{\omega_t} \right)
\Biggl[ \ln^2 \omega_t  + 6 \zeta_2 \Biggr]
\Biggr \}
\nonumber \\[5mm]   &&
+ \frac{(1+y_Z)^2(1+y_Z^2) (17 + 41 y_Z + 17 y_Z^2)}{18 \omega_t y_Z^3}
\Biggl\{
  3 \Biggl( 2  \Li{3}{y_Z} + \Li{3}{-y_Z} \Biggr)
- 3 \zeta_2 \ln (1+y_Z)
\nonumber \\ && \hspace{10mm}
- 2 \ln y_Z \Biggl( 2 \Li{2}{y_Z} + \Li{2}{-y_Z} \Biggr)
- \ln^2 y_Z \Biggl( \ln (1-y_Z) + \frac{1}{2} \ln (1+y_Z) \Biggr)
\Biggr\}
\nonumber \\ &&
+  \frac{(1-y_Z) (1+y_Z)^3 (17 + 41 y_Z + 17 y_Z^2)  }{18 \omega_t y_Z^3}
\times
\nonumber \\ &&\hspace{10mm}
\Biggl\{
  2 \ln y_Z
\Biggl( \ln (1-y_Z) + \frac{1}{2} \ln (1+y_Z)  \Biggr)
+ 2 \Li{2}{y_Z}
+ \Li{2}{-y_Z}
\Biggr\}
\nonumber \\[5mm] &&
- \frac{4}{9} \frac{(1+y_Z)}{y_Z^2}
\left( 5 - 4 \frac{\omega_t y_Z}{(1+y_Z)^2} \right) \times
\nonumber \\ &&
\Biggl\{
\frac{(1+y_Z^2) (1 + 4 y_Z + y_Z^2)}{(1+y_Z)}
\Biggl[
  3 \Biggl( 2 \Li{3}{y_Z} + \Li{3}{-y_Z} \Biggr)
- 3 \zeta_2 \ln (1+y_Z)
\nonumber \\ && \hspace{10mm}
- 2 \ln y_Z \Biggl( 2 \Li{2}{y_Z} + \Li{2}{-y_Z} \Biggr)
- \ln^2 y_Z \Biggl( \ln (1-y_Z) +  \frac{1}{2} \ln (1+y_Z)  \Biggr)
\Biggr]
\nonumber \\ &&
+ (1-y_Z) (1 + 4 y_Z + y_Z^2)
\Biggl[
   2 \ln y_Z \Biggl( \ln (1-y_Z) +  \frac{1}{2} \ln (1+y_Z)  \Biggr)
+  2 \Li{2}{y_Z}
\Biggr]
\nonumber \\ &&
+ \frac{9}{4} (1-y_Z)^2(1+y_Z) \ln (1+y_Z)
+ \frac{\left( 1 + 2 y_Z -24 y_Z^2 + 2 y_Z^3 + y_Z^4 \right)}{(1-y_Z)} \Li{2}{-y_Z}
\nonumber \\ &&
+ \frac{3}{4} \frac{y_Z (1+ 3 y_Z)( 4 - y_Z + y_Z^2)}{(1-y_Z)} \ln y_Z
- \frac{1}{4} \frac{y_Z(2 + 9 y_Z + 3 y_Z^2 + 16 y_Z^3 + 6 y_Z^4)}{(1+y_Z)(1-y_Z)} \ln^2 y_Z
\Biggr\}
\nonumber \\[5mm] &&
- \frac{(1+y_Z)^3 (9 + 32 y_Z + 9 y_Z^2)}{4 \omega_t (1-y_Z) y_Z^2 } \Li{2}{-y_Z}
+ \frac{447}{16} + \frac{125}{9} \frac{(1+y_Z^2)}{y_Z}
\nonumber \\ &&
+ \frac{32}{3} \Biggl[1 - \omega_t \frac{y_Z}{(1+y_Z)^2} \Biggr] \Biggl\{ \zeta_3 - 4  \zeta_2 \ln 2   \Biggr\}
+ \frac{(1+y_Z)^4 (17 y_Z^2 - 19 y_Z + 17)}{8 \omega_t y_Z^3} \ln  (1+y_Z)
\nonumber \\ &&
- \frac{1}{\omega_t} \ln^2 y_Z
\Biggl\{
- \frac{685}{36}
+ \frac{17}{36 y_Z^2}
+ \frac{67}{24 y_Z}
- \frac{335}{24} y_Z
- \frac{497}{72} y_Z^2
- \frac{17}{12} y_Z^3
+ \frac{25}{1-y_Z}
\Biggr\}
\nonumber \\ &&
+ \frac{(1+y_Z)}{24 \omega_t y_Z^2 (1-y_Z)}
\Biggl[
  51  y_Z^5
+ 113 y_Z^4
+ 134 y_Z^3
+ 237 y_Z^2
+ 197 y_Z
+ 68
\Biggr] \ln y_Z
\nonumber \\ &&
- \frac{1}{3} \zeta_2 \Biggl\{
          \frac{20 - 39 y_Z - y_Z^2 - 80 y_Z^3 - 20 y_Z^4}{y_Z(1-y_Z)}
- \omega_t \frac{7 - 49 y_Z + 17y_Z^2 - 55 y_Z^3 - 16 y_Z^4)}{(1-y_Z) (1+y_Z)^2}
\Biggr\}
\nonumber \\ &&
- \frac{1}{\omega_t}
\Biggl\{
\frac{4157}{72}
+ \frac{425 (1+y_Z^4)}{72 y_Z^2}
+ \frac{2561 (1+y_Z^2)}{96 y_Z}
\Biggr\}
\nonumber \\ &&
+ \frac{1}{\omega_t} \zeta_2
\Biggl\{
  \frac{187}{3}
+ \frac{17}{6 y_Z^2}
+ \frac{133}{12 y_Z}
+ \frac{535}{12} y_Z
+ \frac{211}{12} y_Z^2
+ \frac{17}{6} y_Z^3
- \frac{50}{1-y_Z}
\Biggr\}
\nonumber \\ &&
\nonumber \\[5mm]  &&
- 3 \omega_t  \ln \omega_t \frac{(1 + y_H + y_H^2)}{(1+y_H)^2}
+ \frac{3}{2 \omega_t}  \frac{y_H}{(1+y_H)^2}
 \frac{(1+y_Z)^4 }{y_Z^2}  \ln  \frac{(1+y_Z)^2}{y_Z}
\nonumber \\ &&
- \frac{1}{\omega_t}  \frac{y_H}{(1+y_H)^2}
\Biggl\{
  \frac{11 (1+y_H^2) (1+y_H)^2 }{8 y_H^2}
+ 8 N_c
+ \frac{1}{2} \frac{(1+y_Z)^4}{y_Z^2}
\Biggr\}
\nonumber \\ &&
+ \frac{1}{\omega_t} \zeta_2
\Biggl\{
 \frac{3}{2  y_H}
+ \frac{9}{2} y_H
+ \frac{3}{4} y_H^2
\Biggr\}
- \frac{\omega_t }{36} \frac{(625 + 1286 y_H + 625 y_H^2)}{(1+y_H)^2}
\nonumber \\  &&
+ \frac{1}{\omega_t} \frac{(1-y_H)^2}{y_H^2} \ln y_H
\Biggl[ \ln (1-y_H) + \frac{1}{2} \ln (1+y_H) \Biggr]
\Biggl[ (1-y_H^2) - \frac{1}{2} (1+y_H^2) \ln y_H \Biggr]
\nonumber \\ &&
- \frac{1}{8} \frac{1}{\omega_t} \frac{2+8 y_H - 10 y_H^2 - 3y_H^3}{y_H } \ln^2 y_H
+ \frac{1}{8} \frac{1}{\omega_t} \frac{(1+y_H )(6 - 63 y_H + 5 y_H^2)}{y_H } \ln y_H
\nonumber \\ &&
- \frac{1}{8} \frac{1}{\omega_t} \frac{(1+y_H)^2 (5 - 62 y_H + 5 y_H^2)}{y_H^2} \ln (1 + y_H)
- \frac{3}{2} \frac{1}{\omega_t}\zeta_2 \ln (1+y_H) \frac{(1-y_H)^2 (1+y_H^2)}{y_H^2}
\nonumber \\ &&
+ \frac{1}{\omega_t} \frac{(1-y_H) (1+y_H) }{y_H^2}
\Biggl\{
\frac{(5 - 28 y_H + 5 y_H^2)}{4} \Li{2}{-y_H} + (1-y_H)^2 \Li{2}{y_H}
\Biggr\}
\nonumber \\ &&
+ \frac{1}{\omega_t} \frac{(1-y_H)^2 (1+y_H^2)}{y_H^2}
\Biggl\{
\frac{3}{2} \Biggl[2 \Li{3}{ y_H}  + \Li{3}{-y_H}  \Biggr]
- \ln y_H \Biggl[ 2 \Li{2}{y_H} + \Li{2}{-y_H}  \Biggr]
\Biggr\} \Biggr)
\; ,
\nonumber \\ &&
\label{MS2:subtracted}
\end{eqnarray}
where
\begin{eqnarray}
\omega_t & = & \frac{m_W^2}{m_t^2} \; ,
\nonumber \\
y_A & = & \frac{1-\sqrt{1-\frac{4 m_t^2}{m_A^2}}} {1+\sqrt{1-\frac{4 m_t^2}{m_A^2}}} \; , A=H,Z\;\;.
\end{eqnarray}
We have explicitly factorized the RG logarithms,
$C^{(2,1)}_{\alpha \alpha_s} $ and $C^{(2,2)}_{\alpha \alpha_s}$,
which may be calculated also from the one-loop result and
the knowledge of the mass anomalous dimensions (see~\cite{expansion} for
the general case):
\begin{eqnarray}
C^{(2,2)}_{\alpha \alpha_s} & = & Z^{(1,1)}_{\alpha_s}
\left(1 + m_t^2\frac{\partial}{\partial m_t^2} \right) Z^{(1,1)}_{\alpha}
= Z^{(2,2)}_{\alpha \alpha_s}
\\
C^{(2,1)}_{\alpha \alpha_s} & = &
2 Z^{(2,1)}_{\alpha \alpha_s}
+ 4 Z^{(1,1)}_{\alpha} Z^{(1,1)}_{\alpha_s}
+ Z^{(1,1)}_{\alpha} \Delta X^{(1)}_{\alpha_s}
+ Z^{(1,1)}_{\alpha_s}
\left( 1 + 2 m_t^2 \frac{\partial}{\partial m_t^2} \right)
\Delta  X^{(1)}_{\alpha} \;,
\end{eqnarray}
where
\begin{eqnarray}
&&
m_t^2 \frac{\partial}{\partial m_t^2} \Delta  X^{(1)}_{\alpha}  =
   \frac{(1+y_H)^4}{8 \omega_t  y_H^2} \ln(1 + y_H)
-  \frac{(1+y_H) (3+y_H) }{8 \omega_t} \ln y_H
\nonumber \\ &&
+ \frac{1}{\omega_t} \Biggl\{
  \frac{3}{4} \frac{(1+y_Z)^4}{y_Z^2} \frac{y_H}{(1+y_H)^2}
+ \frac{1}{4} \frac{(y_H^2 + 3 y_H + 1)}{y_H}
- \frac{2 y_H}{(1+y_H)^2} N_c
\Biggr\}
\nonumber \\ &&
+ \omega_t \Biggl[ \frac{3}{2} \frac{y_H}{(1+y_H)^2}  - \frac{25}{18} \Biggr]
+ \frac{1}{\omega_t} \Biggl\{
  \frac{(1+y_Z)^2}{3 y_Z}
- \frac{19}{4}
- \frac{17 ( 1+ y_Z^4) }{ 36 y_Z^2}
- \frac{20 (1+ y_Z^2) }{ 9 y_Z}
\Biggr\}
\nonumber \\ &&
- \frac{2}{9}
\left( 5 - 4 \frac{\omega_t y_Z}{(1+y_Z)^2} \right)
\Biggl\{ \frac{(1+y_Z)^4}{y_Z^2} \ln(1 + y_Z)
+ \frac{ y_Z(1+y_Z) (2 + y_Z)}{(1-y_Z)} \ln y_Z
\Biggr\}
\nonumber \\ &&
+ \frac{(1+y_Z)^3}{72 \omega_t}
\Biggl[
\frac{(1+y_Z)}{y_Z^3} (34 + 41 y_Z + 34 y_Z^2) \ln(1 + y_Z)
+ \frac{(41 + 34 y_Z)}{(1-y_Z)} \ln y_Z
\Biggr]
\nonumber \\ &&
+ \frac{169}{72}
+ \frac{10 (1+y_Z^2)}{ 9 y_Z}
+ \frac{1}{8}  \ln \left(1 - \frac{1}{\omega_t} \right)
   \frac{(1-\omega_t) (4 \omega_t^2 + \omega_t + 1)}{\omega_t}
+ \frac{1}{8 \omega_t}  \ln \frac{1}{\omega_t} \;.
\end{eqnarray}
The $C^{(i,j)}$'s in the SM (in contrast to QCD) have non-polynomial
structure in the dimensionless coupling constants which originates
from the tadpole contributions.

\section{Results}

The relation between the top--propagator pole $\tilde{M}$ and the \MSb
mass $m_t$ can be written as

\begin{equation}
\frac{\tilde{M}}{m_t} = 1 + \Biggl \{ \Sigma_1 \Biggr\}_{\overline{\rm MS}}
                   + \Biggl \{ \Sigma_2 + \Sigma_1 \Sigma_1^{'}
\Biggr\}_{\overline{\rm MS}}
+ {\cal O} (g^4,\alpha_s^2)
\;,
\label{result}
\end{equation}
where the r.h.s. is given by Eqs.~(\ref{MS1:subtracted})
and~(\ref{MS2:subtracted}).

We would like to elucidate several aspects of our calculation: all
diagrams have been calculated in the so-called modified $\overline{\rm
MS}$ scheme ($\overline{\rm MMS}$)~\cite{ON3A}, which is defined by
multiplying each loop by the factor
$1/(4\pi)^\ep/\Gamma(1+\varepsilon)$. It has been shown
in~\cite{red,poleI,poleII} that at the two-loop level this scheme is
equivalent to the \MSb scheme when applied to calculating mass
relations. We once more emphasize that the inclusion of the tadpole
diagrams, shown in Fig.~\ref{tadpole}, is important for several
reasons: i) to restore gauge invariance; ii) to get UV counter-terms
which coincide with the ones calculated in the unbroken phase; iii) to
restore the proper form of the RG equations in the broken phase (the
result (\ref{result}) is manifestly RG invariant through ${\cal O}
(g^4,\alpha_s^2)$).

As the top is an unstable particle the pole of the top--propagator
exhibits a real part which is the pole--mass $M_t$ and an imaginary
part which up to a sign gives the width $\Gamma_t$ divided by two (see
(\ref{properMG}) for the precise definitions). The
imaginary part is coming from diagrams with $W,\phi^\pm$ lines (see
Fig.~\ref{diagrams2}) and may be calculated analytically. We find
\begin{eqnarray}
\frac{\Gamma_t}{m_t} &=&-2\:{\rm Im}~\frac{\tilde{M}}{m_t}  =
  \frac{e^2}{8\pi \sin^2 \theta_W}
\Biggl\{ \frac{1}{8 \omega_t} - \frac{3}{8} \omega_t 
+ \frac{1}{4} \omega_t^2 \Biggr\}
\nonumber \\ & + &
 \frac{\alpha_s}{4 \pi} \frac{e^2}{8\pi \sin^2 \theta_W} C_f
\Biggl\{
 \frac{9}{8} \ln \frac{m_t^2}{\mu^2} \left( 2 \omega_t^2 
- \omega_t - \frac{1}{\omega_t}
                                   \right)
+ \frac{1}{8} (1-\omega_t) \omega_t \left( \frac{17}{\omega_t^2} 
+ \frac{21}{\omega_t} + 18 \right)
\nonumber \\ && \hspace{5mm}
- \frac{1}{4}  \ln (1-\omega_t) (1-\omega_t)^2 \left( \frac{5}{\omega_t} 
+ 4 \right)
- \frac{1}{2} \ln \omega_t \left( 1- \omega_t - 2 \omega_t^2 \right)
\nonumber \\ && \hspace{5mm}
-\frac{1}{2}  (1-\omega_t)^2 \left( 2 + \frac{1}{\omega_t} \right)
\Biggl[
\ln \omega_t \ln (1-\omega_t) + 2 \zeta_2 + 2 \Li{2}{\omega_t}
\Biggr]
\Biggr\} 
\label{width_MS}
\end{eqnarray}
as a result in terms of \MSb parameters. With the help of
(\ref{reverse}) we may get the corresponding result in terms of
on--shell parameters
\begin{eqnarray}
\frac{\Gamma_t}{M_t} & = &  \frac{\alpha}{2 \sin^2 \theta_W^{OS}}
\frac{1}{8}
\left( 1- \frac{M_W^2}{M_t^2} \right)^2 \left( 1 
+ 2 \frac{M_W^2}{M_t^2}\right) \frac{M_t^2}{M_W^2}
\nonumber \\ & - &
\frac{\alpha_s}{4 \pi} \frac{\alpha}{2 \sin^2 \theta_W^{OS}}  C_f
\Biggl\{ \frac{1}{8} \left( 1
- \frac{M_W^2}{M_t^2} \right) \frac{M_t^2}{M_W^2}
\left( 6 \frac{M_W^4}{M_t^4} - 9  \frac{M_W^2}{M_t^2}  - 5 \right)
\nonumber \\ && \hspace{5mm}
+ \frac{1}{4}  \ln \left( 1- \frac{M_W^2}{M_t^2} \right)
                  \left( 1- \frac{M_W^2}{M_t^2} \right)^2
                   \left( \frac{5 M_t^2}{M_W^2} + 4 \right)
+ \frac{1}{2} \ln \frac{M_W^2}{M_t^2}
             \left( 1- \frac{M_W^2}{M_t^2} 
- 2  \frac{M_W^4}{M_t^4} \right)
\nonumber \\ && \hspace{5mm}
+ \frac{1}{2} \left( 1- \frac{M_W^2}{M_t^2} \right)^2  \left( 2 
+ \frac{M_t^2}{M_W^2} \right)
\Biggl[
\ln \frac{M_W^2}{M_t^2}  \ln \left( 1- \frac{M_W^2}{M_t^2} \right)  
+ 2 \zeta_2 + 2 \Li{2}{\frac{M^2_W}{M_t^2}}
\Biggr]
\Biggr\} \;,
\nonumber \\
\label{width_OS}
\end{eqnarray}
which coincides with the well know result~\cite{QCD_top_decay} for the
tree and the ${\cal O} (\alpha_s)$ correction to the partial decay
width for the process $t \to b W$ in the approximation of a massless
b-quark.

Very often the inverse of the relation (\ref{result}) is required.  To
that end we have to solve the real part of (\ref{polemass})
iteratively for $m_t$ and to express all $\overline{\rm MS}$
parameters in terms of on-shell ones. The solution to two loops reads
\begin{eqnarray}
\frac{m_t}{M_t } &=& 1 
- {\rm Re} \Biggl \{ \Sigma_1 \Biggr\}_{\overline{\rm MS}}
- {\rm Re} \Biggl \{ \Sigma_2 + \Sigma_1 \Sigma_1^{'}
\Biggr\}_{\overline{\rm MS}}
+ \Biggl[ {\rm Re} \Biggl \{ \Sigma_1 \Biggr\}_{\overline{\rm MS}} \Biggr]^2
\nonumber\\
&&
- \sum\limits_j (\Delta m^2_j)^{(1)} \frac{\partial}{\partial m_j^2} {\rm Re}
\Biggl \{ \Sigma_1 \Biggr\}_{\overline{\rm MS}}
       - (\Delta e)^{(1)} \frac{\partial}{\partial e} {\rm Re}
\Biggl \{ \Sigma_1 \Biggr\}_{\overline{\rm MS}}
      \Biggr|_{m_j^2=M_j^2,\, e=e_{\rm OS}},
\label{reverse}
\end{eqnarray}
where the sum runs over all species of particles $j=Z,\,W,\,H, \, t$
and $$ (\Delta m^2_j)^{(1)}= - M_j^2 \frac{e^2_{\rm OS}}{16 \pi^2
\sin^2 \theta_W} X_j^{(1)} \Biggr|_{m_j^2=M_j^2} $$ stands for the
self-energy of the $j$th particle at $p^2=-m_j^2$ in the \MSb scheme
and parameters replaced by the on-shell ones.  The values of
$X_j^{(1)}$ are given in Appendix B of \cite{poleI,poleII}.  The
relation (\ref{reverse}) includes also the transition from the
\MSb to on-shell scheme for the electric charge.  It can be described
as $$ (\Delta e)^{(1)} \frac{\partial}{\partial e} {\rm Re} \Biggl \{
\Sigma_1 \Biggr\}_{\overline{\rm MS}} =
\left\{
  \delta \alpha_{\rm bos}
\!+\! \delta \alpha_{\rm lep}
\!+\! \delta \alpha_{\rm top}
\!+\! \Delta \alpha _{\rm hadrons}^{(5)}(M^2_Z)
\!-\! \delta \Delta \alpha _{\rm udscb}(M^2_Z)\right\}
{\rm Re} \Biggl \{ \Sigma_1 \Biggr\}_{\overline{\rm MS}} \;, $$ where
$$ \delta \alpha_{\rm bos} = \frac{\alpha}{4\pi} \left(
7\ln\frac{M_W^2}{\mu^2}-\frac{2}{3} \right),\;\; \delta \alpha_{\rm
lep} = - \frac{\alpha}{3\pi}
\sum\limits_{\ell=e,\mu,\tau}\ln \frac{m^2_\ell}{\mu^2},\;\;
 \delta \alpha_{\rm top} =  - \frac{4 \alpha}{9 \pi} \ln \frac{m^2_t}{\mu^2}$$
and  $$\delta \Delta \alpha _{\rm udscb}(M^2_Z)=
\frac{11\alpha}{9\pi} \left(\ln\frac{M_Z^2}{\mu^2}-\frac{5}{3} \right) \;.$$
For numerical estimations of $\Delta \alpha _{\rm
hadrons}^{(5)}(M^2_Z)$ we use the results of \cite{FJ03}
\begin{equation}
\Delta \alpha _{\rm hadrons}^{(5)}(M^2_Z) = 0.027773 \pm 0.000354\;\;;\;\;\;
\alpha^{-1}(M_Z^2)=128.922 \pm 0.049
\end{equation}
at $M_Z=$ 91.19 GeV.

\begin{figure}[ht]
\begin{picture}(120,60)(-5,0)
\includegraphics[scale=0.55]{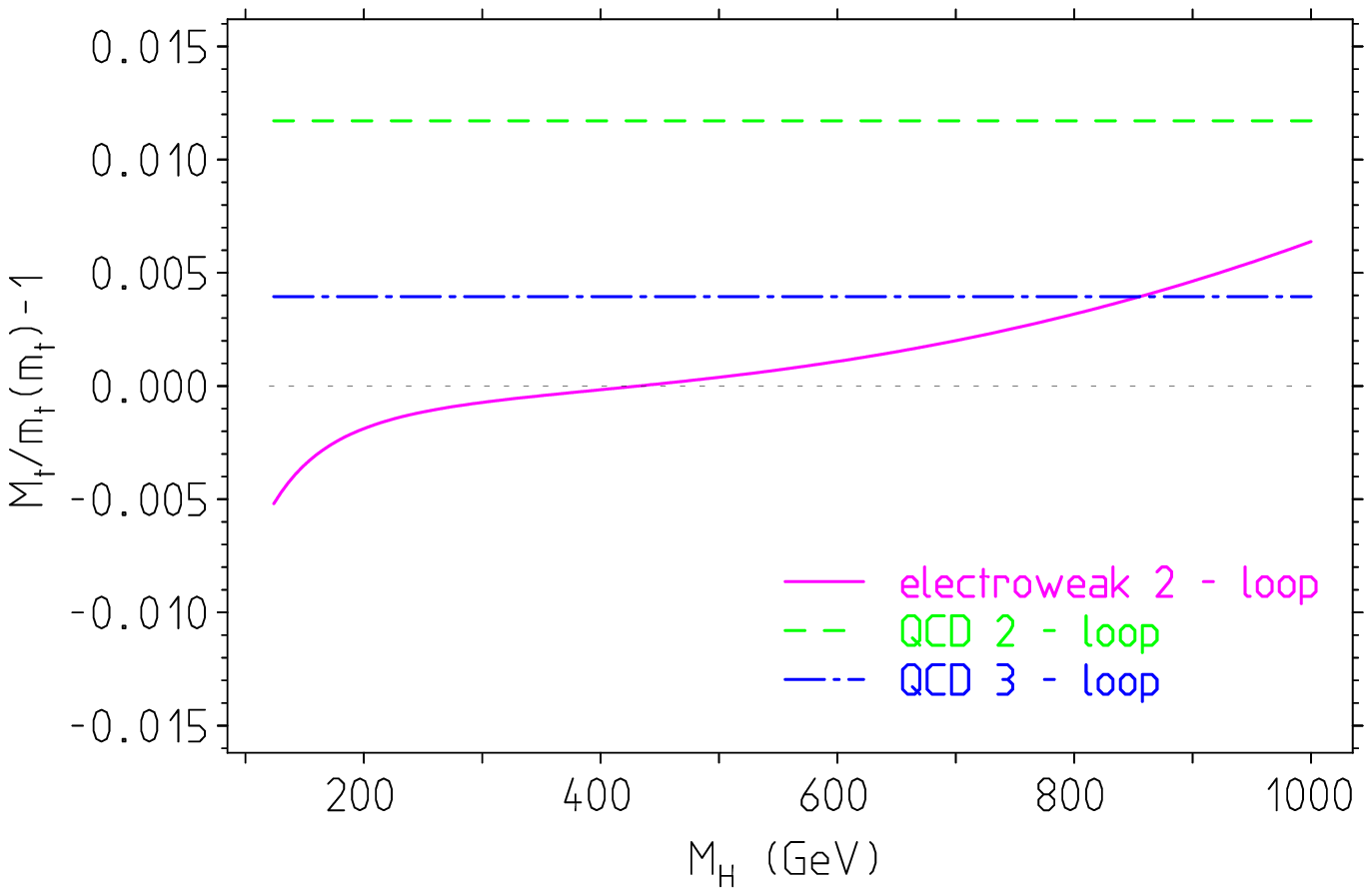}
\end{picture}

\vspace*{-2.2cm}

\begin{picture}(120,60)(-240,0)
\includegraphics[scale=0.55]{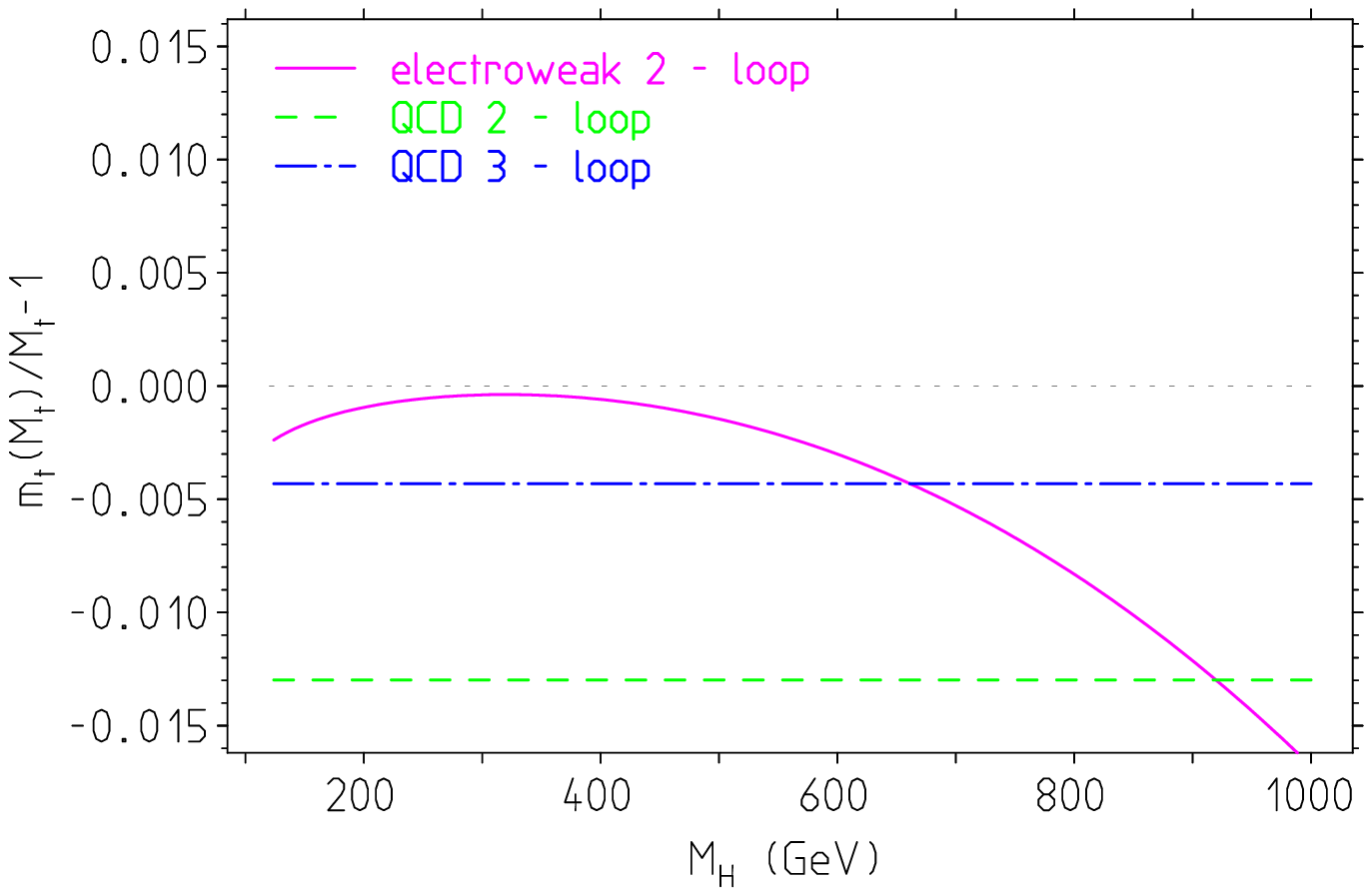}
\end{picture}


\caption[]{Electroweak $O(\alpha \alpha_s)$ correction to 
$M_t/m_t(m_t)-1$ [left] and $m_t(M_t)/M_t-1$ [right], in comparison with
$O(\alpha_s^2)$ and $O(\alpha_s^3)$ QCD corrections as a function of the
Higgs boson mass $M_H$.}
\label{fig:plots}
\end{figure}

For the ${\cal O}(\alpha \alpha_s)$ contribution only the expression
(\ref{reverse}) actually simplifies to
\begin{eqnarray}
\frac{m_t}{M_t }
&=& 1 \!-\! {\rm Re} \Biggl \{ \Sigma_1 \Biggr\}_{\overline{\rm MS}}
\!-\! {\rm Re} \Biggl \{ \Sigma_2 \!+\! \Sigma_1 \Sigma_1^{'}
\Biggr\}_{\overline{\rm MS}}
\!+\! 2 \frac{\alpha_s}{4 \pi} \frac{e^2}{16\pi^2 \sin^2 \theta_W^{OS}}
X_{\alpha}^{(1)}  X_{\alpha_s}^{(1)}
\nonumber\\
&&
+ \frac{\alpha_s}{4 \pi} \frac{e^2}{16\pi^2 \sin^2 \theta_W^{OS}}
2 \Biggl\{
  X_{\alpha}^{(1)} Z_{\alpha_s}^{(1,1)}
+ X_{\alpha_s}^{(1)} m_t^2 \frac{\partial}{\partial m_t^2} X_\alpha^{(1)}
\Biggr\},
\label{reverse:a}
\end{eqnarray}
where
\begin{eqnarray}
m_t^2 \frac{\partial}{\partial m_t^2} X_\alpha^{(1)} & = & m_t^2
\frac{\partial}{\partial m_t^2} \Delta X_\alpha^{(1)} + \ln
\frac{m_t^2}{\mu^2}\: m_t^2 \frac{\partial Z_\alpha^{(1,1)} }{\partial
m_t^2} + Z_\alpha^{(1,1)}
\nonumber \\ & = &
m_t^2 \frac{\partial}{\partial m_t^2} \Delta X_\alpha^{(1,1)} +
Z_\alpha^{(1,1)} + \ln \frac{m_t^2}{\mu^2} \left(\frac{3}{8}
\frac{m_t^2}{m_W^2} + 2 N_c \frac{m_t^4}{m_W^2 m_H^2} \right) \;.
\nonumber
\end{eqnarray}
For Eq.(\ref{reverse:a}) we present a semi--numerical result for
$\mu=M_t$ and the following input parameters: $\alpha$ = 1/137.036,
$M_W$ = 80.419 GeV, $M_Z$ = 91.188 GeV and $M_t$ = 174.3 GeV,

\begin{eqnarray}
&& \hspace{-1.5cm}
\frac{m_t(M_t)}{M_t} =
1 - C_f \frac{\alpha_s}{\pi}
  - \frac{\alpha}{4 \pi \sin^2 \theta_W^{OS}}
\Biggl( 0.3747795 +
\frac{1}{2} \frac{M_W^2}{M_H^2} \left[1 - 3 \ln \frac{M_W^2}{M_t^2} \right]
\nonumber \\ && \hspace{20mm} \hspace{-1.5cm}
- \frac{3}{4} \frac{M_Z^4}{M_W^2 M_H^2} \ln \frac{M_Z^2}{M_t^2}
+ \frac{M_t^4}{M_W^2 M_H^2}
  \Biggl\{
 \frac{1}{2} \frac{M_H^2}{M_t^2}  \frac{(1+Y_H^2)}{Y_H}
+ \frac{1}{4} \frac{M_Z^4}{M_t^4}
- 3
  \Biggr\}
\nonumber \\ && \hspace{20mm} \hspace{-1.5cm}
- \frac{1}{8} \frac{M_H^4}{M_t^2 M_W^2} \ln (1+Y_H)
+ \frac{1}{8} \frac{M_H^2}{M_W^2} \frac{(3+Y_H^2)}{(1 + Y_H)} \ln Y_H
\Biggr)
\nonumber \\ && \hspace{-1.5cm}
- C_f \frac{\alpha_s}{4 \pi} \frac{\alpha}{4 \pi \sin^2 \theta_W^{OS}}
\Biggl(
- 78.591
+ 6 \frac{M_W^2}{M_H^2} \ln \frac{M_W^2}{M_t^2}
+ 3  \frac{M_Z^4}{M_W^2 M_H^2} \ln \frac{M_Z^2}{M_t^2}
- 2 \frac{M_W^2}{M_H^2}
\nonumber \\ && \hspace{-1.5cm}
+ \frac{M_t^4}{M_H^2 M_W^2}
\Biggl\{
- \frac{11}{8} \frac{M_H^2}{M_t^2} \frac{(1+Y_H^2)}{Y_H}
- \frac{M_Z^4}{M_t^4}
+ 6 
\Biggr\}
+ \zeta_2  \frac{M_t^2}{M_W^2}
\Biggl\{
 \frac{3}{2  Y_H}
+ \frac{9}{2} Y_H
+ \frac{3}{4} Y_H^2
\Biggr\}
\nonumber \\  && \hspace{-1.5cm}
+ \frac{M_t^2}{M_W^2} \frac{(1-Y_H)^2}{Y_H^2} \ln Y_H
\Biggl[ \ln (1-Y_H) + \frac{1}{2} \ln (1+Y_H) \Biggr]
\Biggl[ (1-Y_H^2) - \frac{1}{2} (1+Y_H^2) \ln Y_H \Biggr]
\nonumber \\ && \hspace{-1.5cm}
- \frac{1}{8} \frac{M_t^2}{M_W^2} \frac{2+8 Y_H - 10 Y_H^2 - 3Y_H^3}{Y_H } \ln^2 Y_H 
+ \frac{1}{8} \frac{M_t^2}{M_W^2} (1+Y_H)(11 Y_H - 39) \ln Y_H
\nonumber \\ && \hspace{-1.5cm}
- \frac{1}{8} \frac{M_H^2}{M_W^2} \frac{(11 - 50 Y_H + 11 Y_H^2)}{Y_H} \ln (1 + Y_H) 
- \frac{3}{2} \frac{M_t^2}{M_W^2}\zeta_2 \ln (1+Y_H) \frac{(1-Y_H)^2 (1+Y_H^2)}{Y_H^2}
\nonumber \\ && \hspace{-1.5cm}
+ \frac{M_t^2}{M_W^2} \frac{(1-Y_H) (1+Y_H) }{Y_H^2}
\Biggl\{
\frac{(5 - 28 Y_H + 5 Y_H^2)}{4} \Li{2}{-Y_H} + (1-Y_H)^2 \Li{2}{Y_H}
\Biggr\}
\nonumber \\ && \hspace{-1.5cm}
+ \frac{M_t^2}{M_W^2} \frac{(1\!-\!Y_H)^2 (1\!+\!Y_H^2)}{Y_H^2}
\Biggl\{
\frac{3}{2} \biggl[2 \Li{3}{ Y_H} \!+\! \Li{3}{-Y_H}  \biggr]
\!-\! \ln Y_H \biggl[ 2 \Li{2}{Y_H} \!+\! \Li{2}{-Y_H}  \biggr]
\Biggr\}
\Biggr),
\label{number}
\end{eqnarray}
where $Y_A$ is defined via the pole masses as $$ Y_A =
\frac{1-\sqrt{1-\frac{4 M_t^2}{M_A^2}}} {1+\sqrt{1-\frac{4
M_t^2}{M_A^2}}} \; , A=H,Z\;\;.  $$ For illustration of the numerical
significance of our result we shown the two-loop ${\cal O}(\alpha
\alpha_s)$ corrections to $M_t/m_t(m_t)$  and
$m_t(M_t)/M_t$ (Fig.~\ref{fig:plots}) as a function of the Higgs boson
mass in comparison with the two- and three-loop QCD ones~\cite{pole3}
(for simplicity, we consider the case $\alpha_s(\mu) = \alpha_s(M_Z)$).
In a wide range of values of Higgs boson mass (100 GeV $<M_H<1000$ GeV)
our result is comparable in size to the 3-loop QCD corrections. The
corrections grow with the Higgs mass, again this is a consequence of the
breakdown of the decoupling of heavy particles in a ``spontaneously
broken'' gauge theory.

\vspace*{3.0cm}

\pagebreak
\section{Conclusion}
The main result of the present investigation is the
two-loop ${\cal O}(\alpha \alpha_s)$ relationship between
pole-- and \MSbn--mass for the top-quark within the SM.
It is given by Eq.~(\ref{MS2:subtracted}).
Numerically its size is comparable to the 3-loop QCD--correction
for a light Higgs boson and it increases for a heavy Higgs boson.
As a byproduct, several new massive master-integrals
have been calculated analytically (Sec.~3).

\vspace{8mm}
{\bf Acknowledgments.}  We are grateful to A.~Davydychev and O.V.~Tarasov 
for useful discussions.  
We would like to thank K.~Chetyrkin for his interest in our work.  
This research was supported in part by the Australian
Research Council grant No.~A00000780
and by RFBR \# 04-02-17149.


\end{document}